\documentclass[preprint,12pt]{elsarticle}

\usepackage{amssymb}

\usepackage{multirow}

\journal{Nuclear Physics B}

\begin{document}

\begin{frontmatter}



\title{Calibration of the LIGO Gravitational Wave Detectors in the Fifth Science Run} 



\author{J.~Abadie$^{19}$, 
B.~P.~Abbott$^{19}$, 
R.~Abbott$^{19}$, 
M,~Abernathy$^{51}$, 
C.~Adams$^{21}$, 
R.~Adhikari$^{19}$, 
P.~Ajith$^{19}$, 
B.~Allen$^{2,78}$, 
G.~Allen$^{37}$, 
E.~Amador~Ceron$^{63}$
% ...
}
% ...
\address{$^{2}$Albert-Einstein-Institut, Max-Planck-Institut f\"ur Gravitationsphysik, D-30167 Hannover, Germany}
% ...
\address{$^{19}$LIGO - California Institute of Technology, Pasadena, CA  91125, USA }
% ...
\address{$^{21}$LIGO - Livingston Observatory, Livingston, LA  70754, USA }
% ...
\address{$^{37}$Stanford University, Stanford, CA  94305, USA }
% ...
\address{$^{51}$University of Glasgow, Glasgow, G12 8QQ, United Kingdom }
% ...
\address{$^{63}$University of Wisconsin--Milwaukee, Milwaukee, WI  53201, USA }
% ...



\begin{abstract}
The Laser Interferometer Gravitational Wave Observatory (LIGO) is a network of three detectors built to detect local perturbations in the space-time metric from astrophysical sources. These detectors, two in Hanford, WA and one in Livingston, LA, are power-recycled Fabry-Perot Michelson interferometers.  In their fifth science run (S5), between November 2005 and October 2007, these detectors accumulated one year of triple coincident data while operating at their designed sensitivity. In this paper, we describe the calibration of the instruments in the S5 data set, including measurement techniques and uncertainty estimation.

\end{abstract}


\end{frontmatter}


\section{Introduction}
\label{intro}
The Laser Interferometer Gravitational Wave Observatory (LIGO) is a network of three detectors built in the United States to detect local perturbations in the space-time metric from astrophysical sources. These distant sources, including binary black hole or neutron star coalescences, asymmetric rapidly spinning neutron stars, and supernovae are expected to produce time-dependent strain $h(t)$ observable by the interferometer array \cite{kip,300years}.  

\begin{figure}[h!]
\begin{center}
\includegraphics[width=90mm]{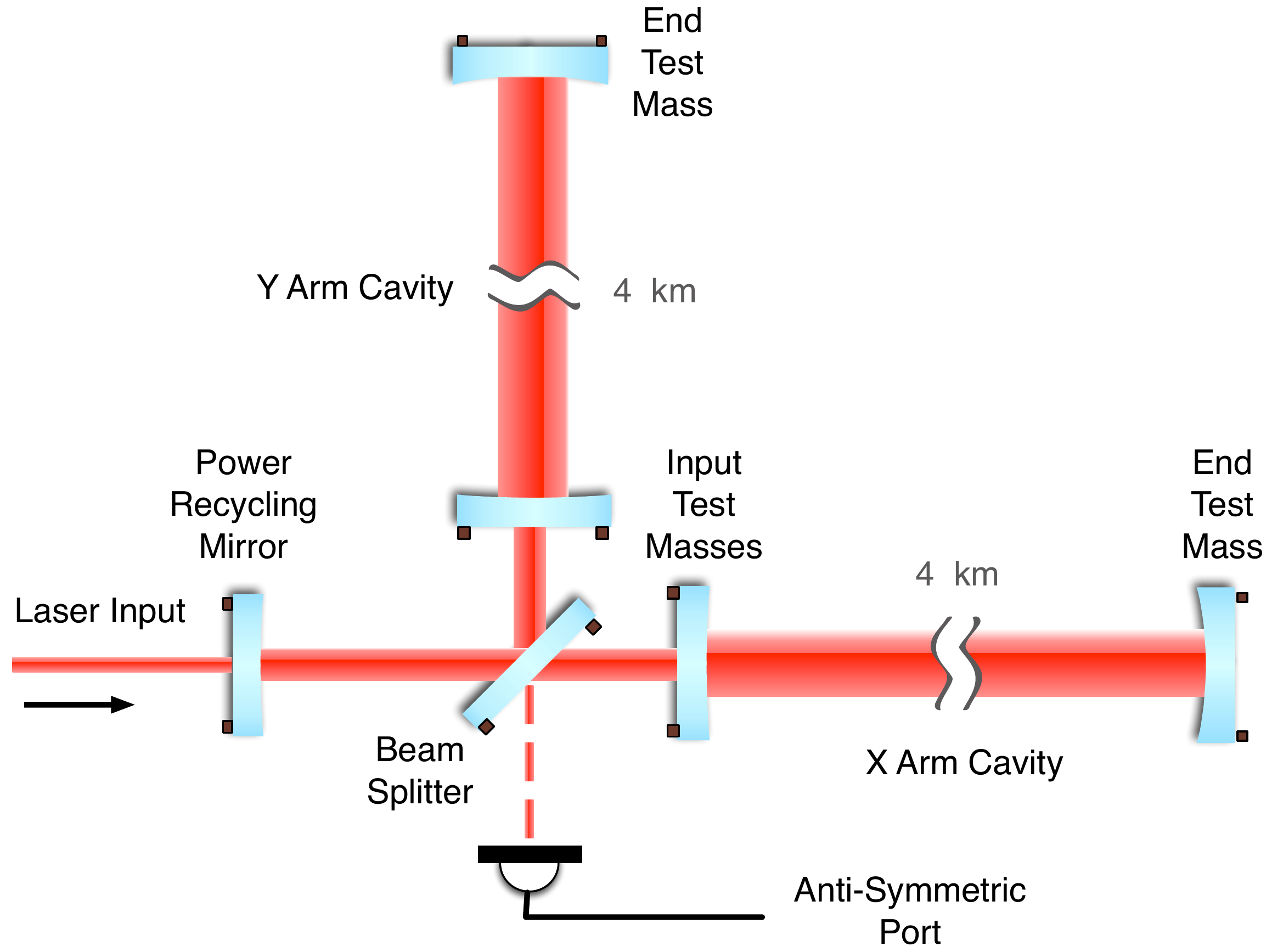}
\end{center}
\caption{Schematic optical layout of the LIGO interferometers. \label{IFOSchematic}}
\end{figure}

\begin{figure}[h!]
\begin{center}
\includegraphics[width=50mm]{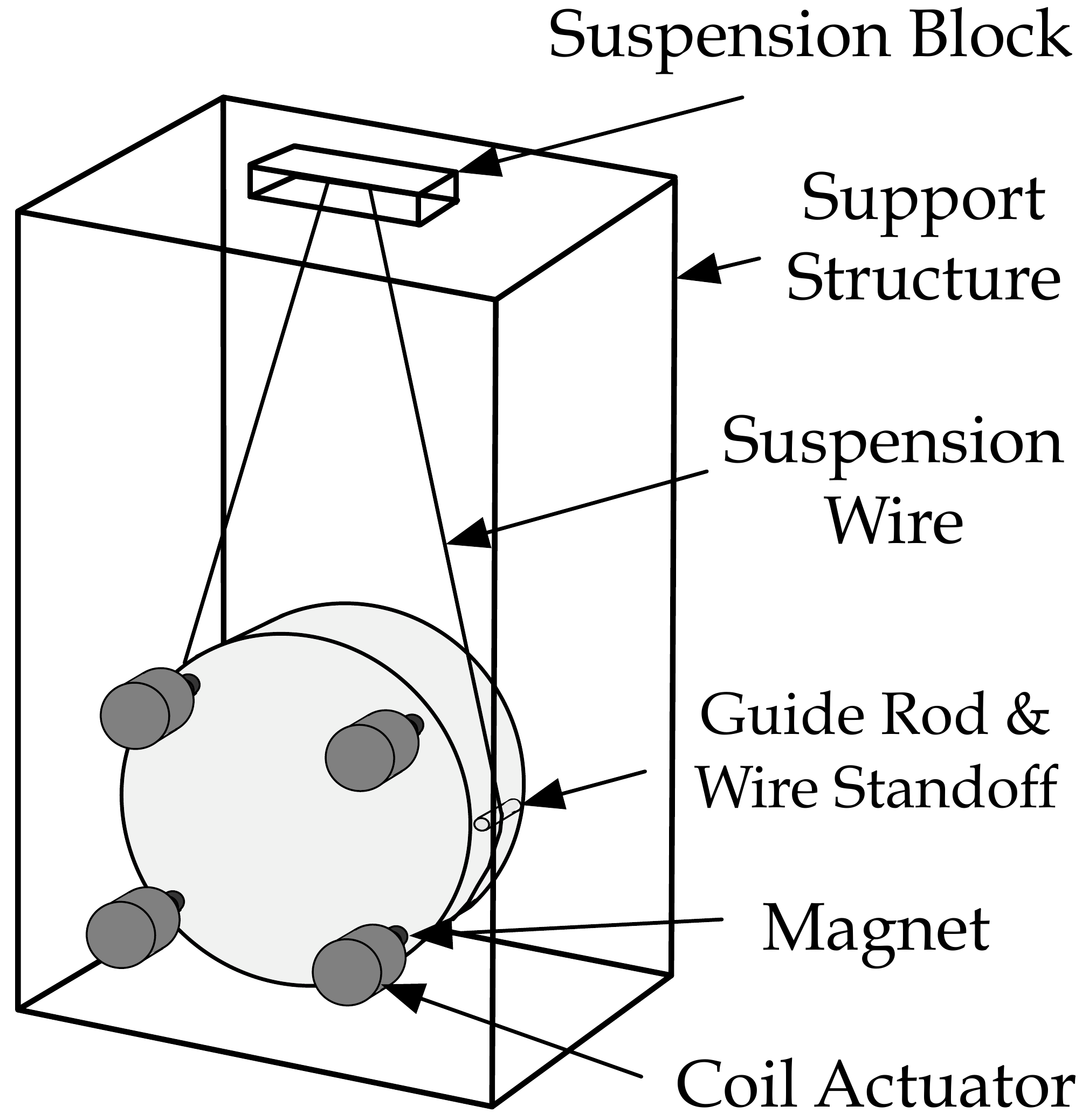}
\caption{A schematic of the LIGO optic suspensions for S5. The actuation force is provided by the coil actuators (mounted to the support structure) which act  upon the magnets secured directly on the rear face of the optic. \label{sus}}
\end{center}
\end{figure}

The detectors, two in Hanford, WA (H1 and H2) and one in Livingston, LA (L1), are power-recycled Fabry-Perot Michelson interferometers. The optical layout of the interferometers is shown in Figure \ref{IFOSchematic}. The perpendicular Fabry-Perot arm cavities of the Michelson, each of length $L=$ 3995 m for H1 and L1 ($L=$ 2009 m for H2), are composed of 10 kg optics or ``test masses'' suspended as pendula. Light reflected from the input port of the Michelson is recycled with an additional suspended optic forming a power recycling cavity. Each interferometer uses a Nd:YAG laser ($\lambda~=~$1064 nm, or $f~=~$282 THz), whose phase is modulated at several frequencies such that a Pound-Drever-Hall style control scheme \cite{pdh1,pdh2} can be used to hold the arm cavities and power recycling cavity in resonance. Figure \ref{sus} shows a schematic of the suspension system for a given optic and electro-magnetic coil-actuators (paired with magnets secured on the rear face of the optic) used to control its motion.  Further details of the interferometer configuration are described in \cite{detectorpaper}.

\begin{figure}[h!]
\begin{center}
\includegraphics[width=90mm]{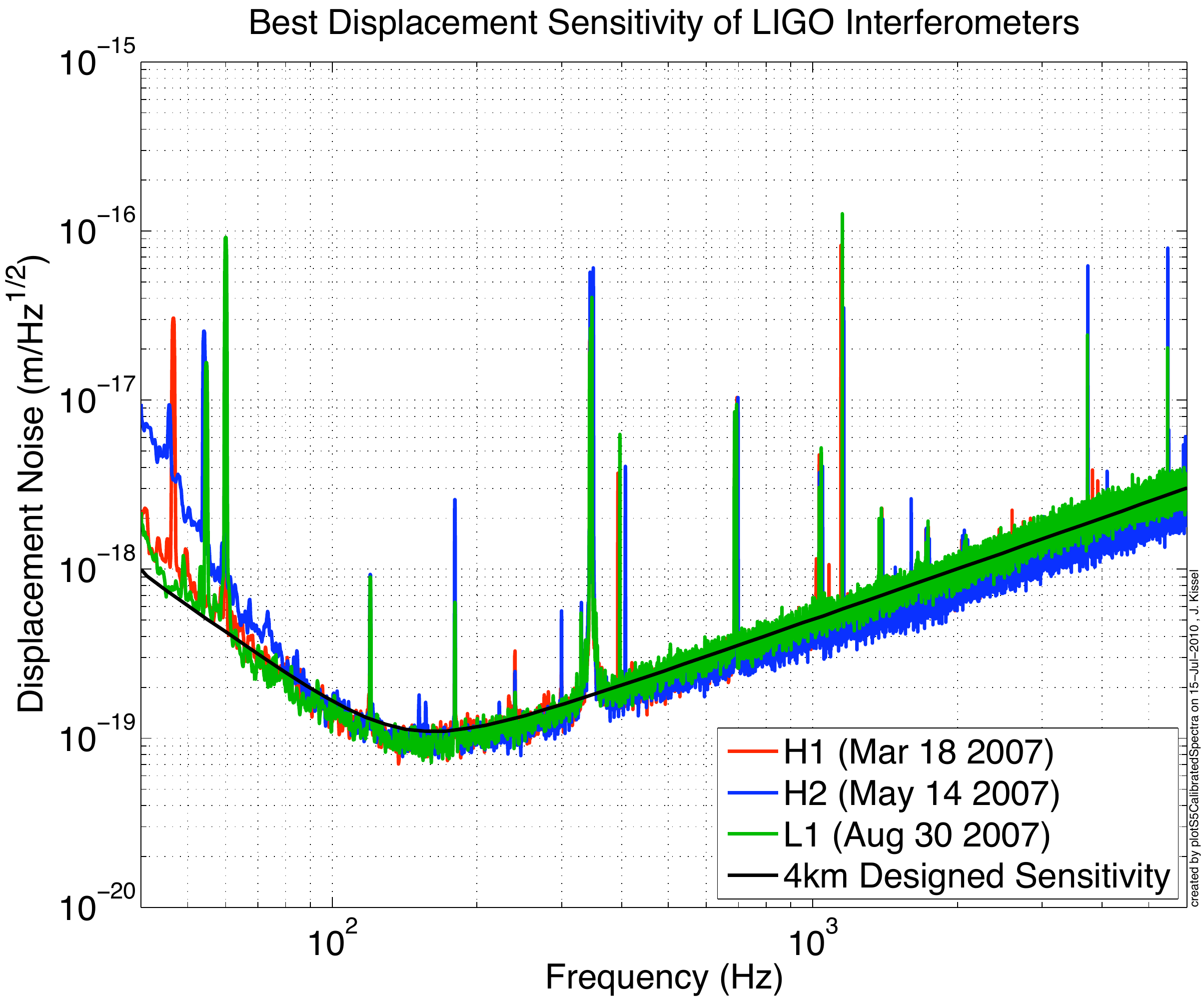}
\end{center}
\caption{The best displacement sensitivity, expressed as equivalent displacement noise, for each interferometer during the S5 science run, and expected total noise in LIGO's first 4 km interferometers. \label{S5Curves}}
\end{figure}

\begin{figure}[h!]
\begin{center}
\includegraphics[width=90mm]{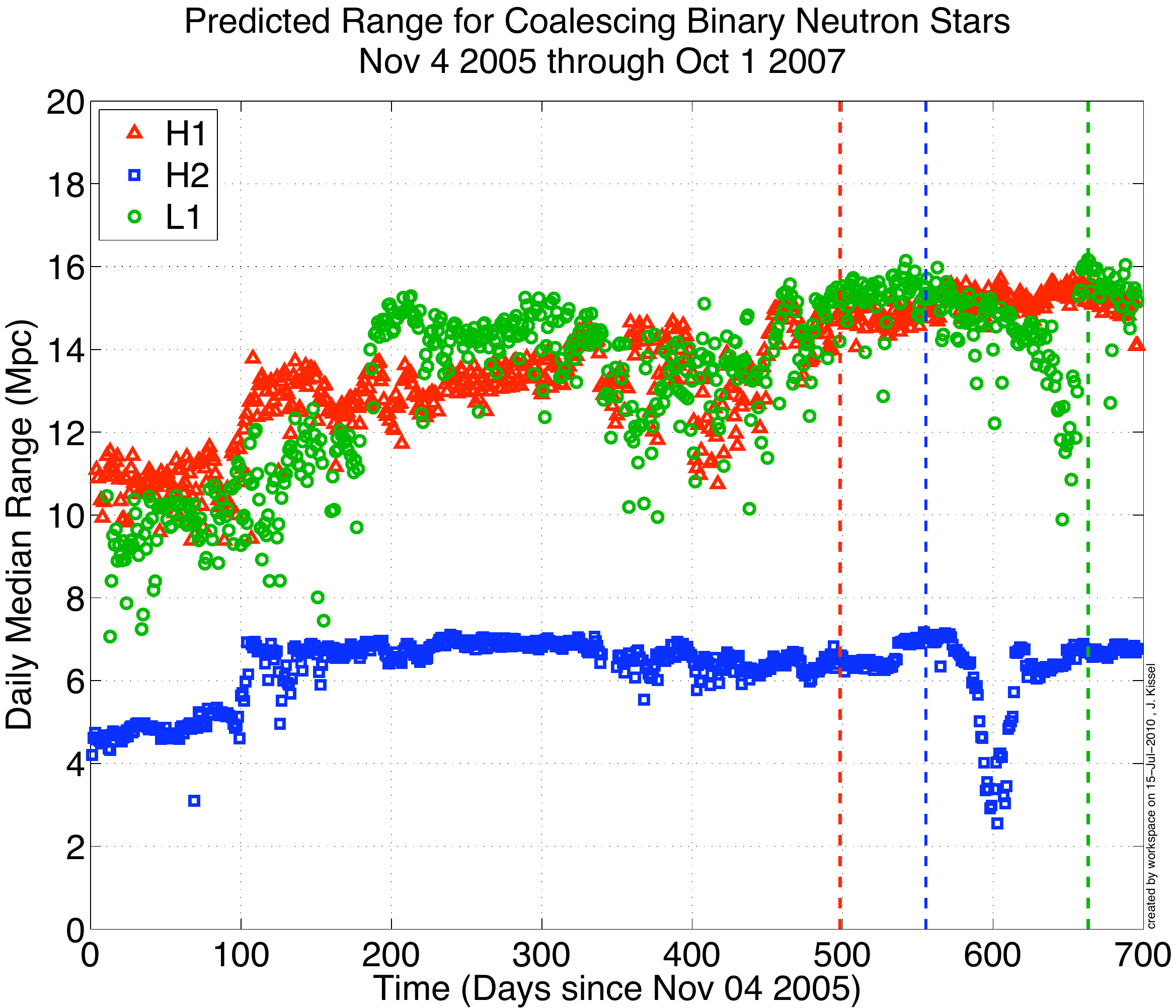}
\end{center}
\caption{Daily median of detector sensitivity during S5 to a 1.4-1.4 solar mass compact binary system averaged over angle and orientation. Dashed lines indicate the times during which the representative spectra in Figure \ref{S5Curves} were taken. Large variations in detector sensitivity are due to upgrades or hardware problems. \label{range}}
\end{figure}

During the fifth LIGO science run (S5), these detectors accumulated approximately one year (368.84 days) of triple coincidence data near their designed sensitivity between Nov 4 2005 and Oct 1 2007 (GPS time 815097613 through 875232014). The best sensitivity (strain amplitude spectral density) for each detector and an example sensitivity curve used to guide the design for the 4 km detectors \cite{srd} are shown in Figure \ref{S5Curves}. As a figure of merit of the sensitivity over time, we integrate the power spectral density using a matched-filter template describing a binary neutron star (1.4-1.4 solar mass) coalescence over which angle and orientation have been averaged. This metric produces a predicted range out to which we may see such a source with signal-to-noise ratio of 8 (see \cite{s5insprange} for details). Figure \ref{range} illustrates the daily median of this range over the course of the science run. 

Differential displacement of the interferometer's end test masses is measured by precisely monitoring the differential phase between light returned by each Fabry-Perot arm cavity using a Pound-Drever-Hall error signal. When the interferometer is under servo control, this error signal $e_{D}(f)$ is proportional to a differential arm (DARM) length change, $\Delta L_{ext}(f)$ caused by the end test mass displacement such that,
\begin{eqnarray}
\Delta L_{ext} (f) = R_{L}(f)~e_{D}(f) \label{lineardarmloop}
\end{eqnarray}
where the change in length $\Delta L_{ext}$ is the sum of the interferometer's response to the astrophysical signal and other differential noise sources. 

The quantity $R_{L}(f)$ is a complex function in the frequency-domain known as the ``length response function.'' In this paper, we provide a complete description of a frequency-domain model of the length response function used for each detector in the S5 data set. Table \ref{errorsummary} summarizes the uncertainty in our model of  $R_{L}(f)$, broken up into magnitude and phase of the complex function, and separated into three frequency bands. Each value is the estimated 68\% confidence interval (one sigma) across the band for the entire 2 calendar-year science run. 

 \begin{table}[h!]
\label{errorsummary}
\caption{Summary of band-limited response function errors for the S5 science run.}
\begin{center}
\begin{tabular}{|c|c|c|c||c|c|c|}
\hline
\multirow{3}{*}{{\bf }} & \multicolumn{3}{|c||}{{\bf $R_{L}(f)$ Magnitude Error (\%)}} & \multicolumn{3}{c|}{{\bf $R_{L}(f)$ Phase Error (Deg)}}\\ 
& &  &  &  &  & \\
 & 40-2000 Hz & 2-4 kHz & 4-6 kHz & 40-2000 Hz & 2-4 kHz & 4-6 kHz \\ \hline
H1 & 10.4 & 15.4 & 24.2  & 4.5 & 4.9 & 5.8 \\
H2 & 10.1 & 11.2 & 16.3  & 3.0 & 1.8 & 2.0 \\
L1 & 14.4 & 13.9 & 13.8 & 4.2 & 3.6 & 3.3 \\
\hline
\end{tabular}
\end{center}
\end{table}

In Section \ref{model}, we describe the model used for all LIGO interferometers which divides a given interferometer into three major subsystems -- sensing, digital control, and actuation -- and includes a detailed description of the important components of each subsystem. Measurements of these components along with corresponding uncertainties are presented in Section \ref{meas}. Finally, the response function, $R_{L}(f)$, is developed from the subsystems and the uncertainty in each subsystem are combined in Section \ref{error} to form the total uncertainty estimate as seen in Table \ref{errorsummary}. 

Gravitational wave data analysis is performed on a signal proportional to strain generated in the time domain from $e_{D}(t)$ and a convolution kernel, $R_{L}(t - t')$,
\begin{eqnarray}
h(t) = \frac{1}{L} \int R_{L}(t - t')~e_{D}(t')~dt',
\end{eqnarray}
developed from the parameters of the length response function. The production of the time-domain convolution kernel, $R_{L}(t-t')$, from the frequency-domain model, $R_{L}(f)$, and the associated additional uncertainty are discussed in detail in \cite{hoft,V3V4hofterrors}. 

\section{Model}
\label{model}

\begin{figure}[h]
\begin{center}
\includegraphics[width=70mm]{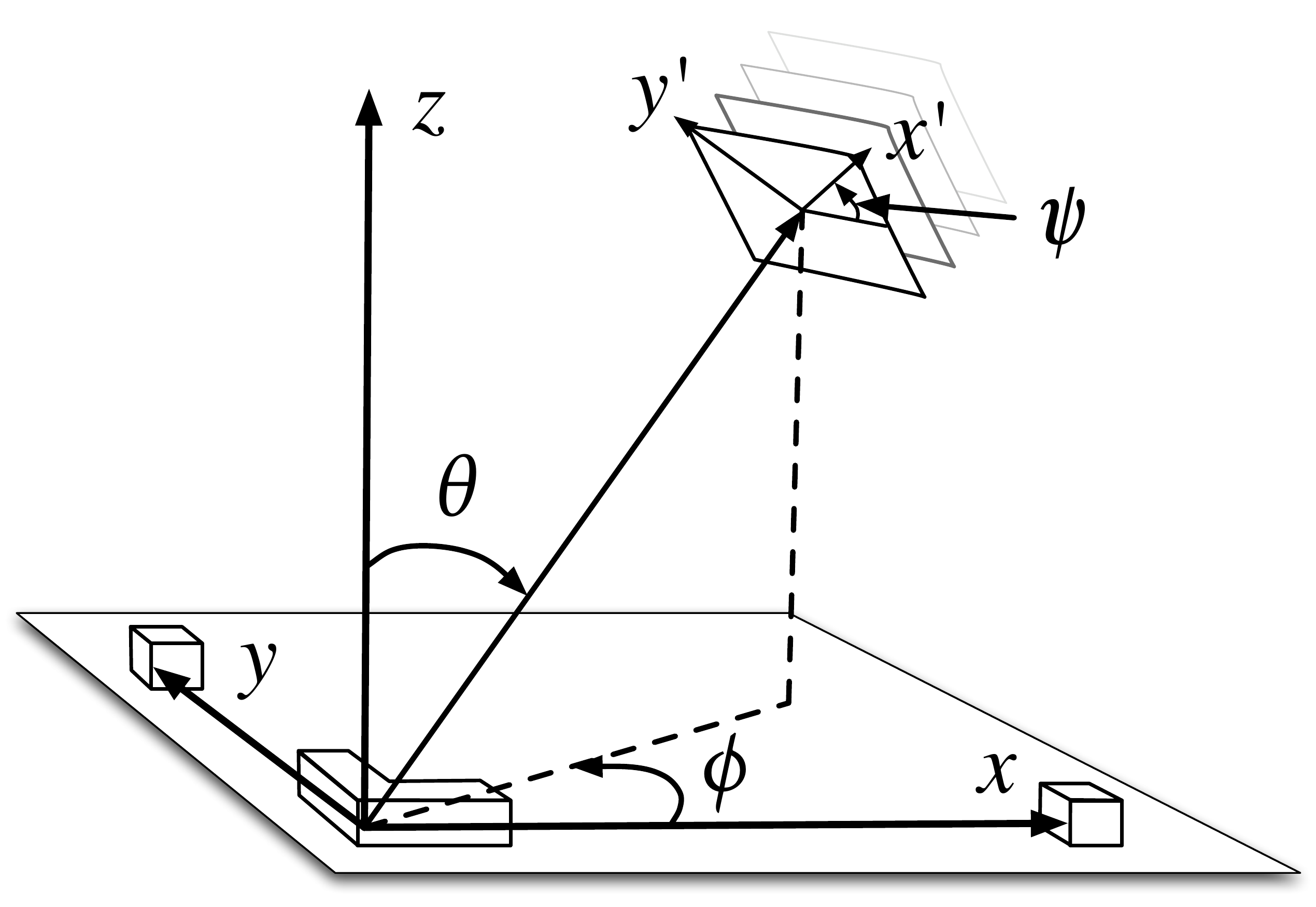}
\end{center}
\caption{A schematic of the coordinates used both in the interferometer basis, and in the incoming plane wave basis. The Euler angles $\theta,\phi$ and $\psi$ are as defined in \cite{excesspower,fplusfcross}, except here they are shown relative to the detector frame rather than the equatorial frame. \label{ifocoords}}
\end{figure}

Astrophysical gravitational wave strain $h(f)$ detected by the interferometers contains source information including wave forms $h_{+,\times}(f)$, azimuthal angle $\phi$, polar angle $\theta$, and orientation (or polarization angle) $\psi$ (see Figure \ref{ifocoords}). The amplitude of the wave's projection into the interferometer basis is described by 
\begin{eqnarray}
h(f) = F_{\times}(\theta, \phi,\psi) h_{\times}(f) + F_{+}(\theta, \phi, \psi) h_{+}(f), \label{cgw}
\end{eqnarray}
where $F_{\times,+}$ are the antenna response of the detectors and $h_{\times,+}$ are the wave amplitudes in the ``cross'' and ``plus'' polarizations of the local metric perturbations $h_{\mu \nu}$ in the transverse-traceless gauge \cite{kip,excesspower,fplusfcross}. 

We model each interferometer's repsonse to an optimally-oriented ($\theta = \phi = \psi = 0$), plus-polarized wave form using the long wavelength approximation. The approximation is valid between 40 and 6000 Hz, and has associated uncertainty of at most 2\% \cite{hfcorrs}. From this reference model, the detector response to an arbitrary waveform, orientation, and polarization angle may be calculated analytically \cite{kip,excesspower,fplusfcross}. In the long wavelength approximation, the strain amplitude, $h(f)$, in Fabry-Perot arm lengths of the interferometer is
\begin{eqnarray}
h(f)  & = & \frac{L^{x}_{ext}(f) - L^{y}_{ext}(f)}{L} = \frac{\Delta L_{ext}}{L}.
\end{eqnarray}

\begin{figure}[h!]
\begin{center}
\includegraphics[width=90mm]{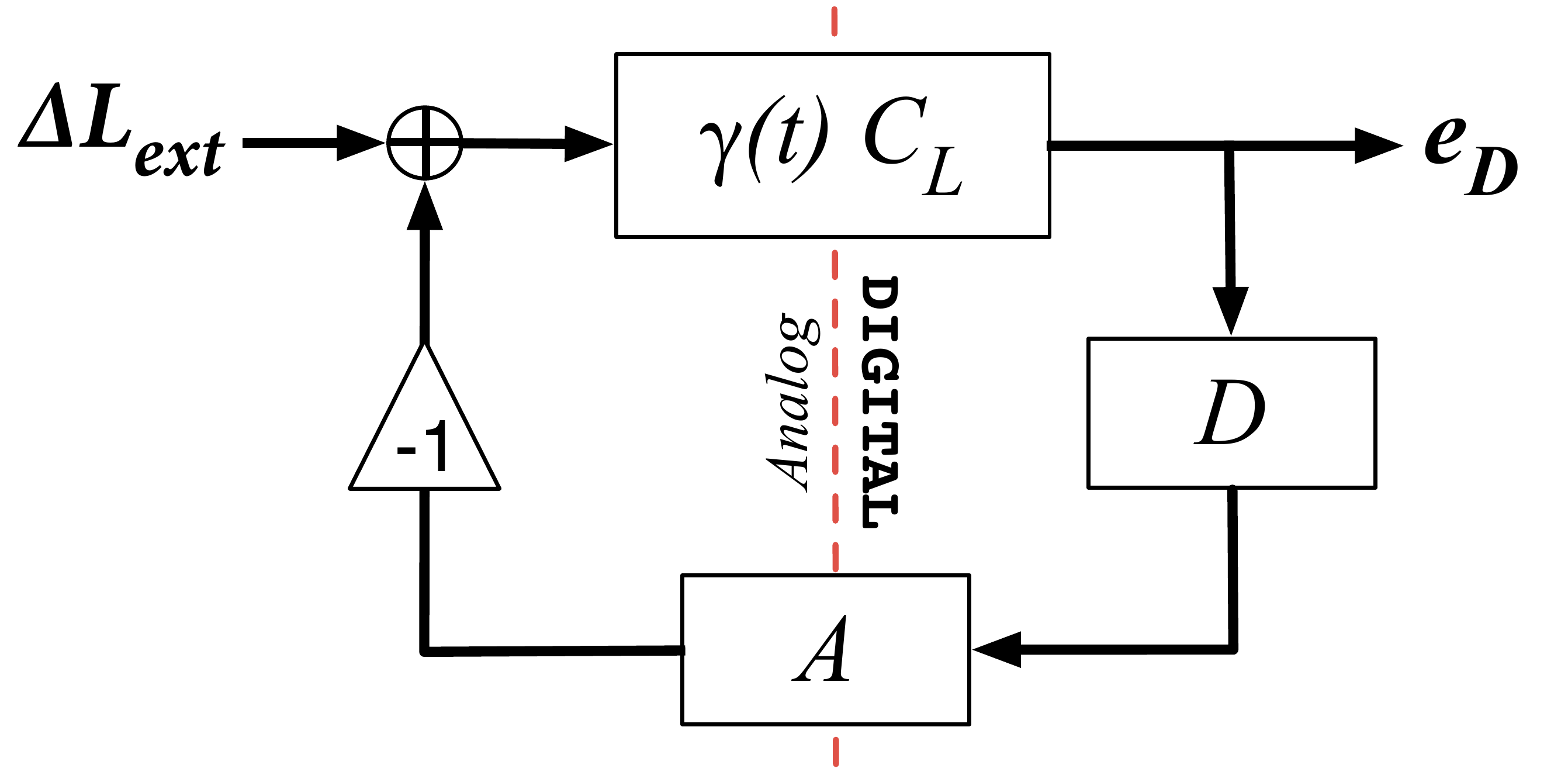}
\end{center}
\caption{The single-input, single-output model of the control loop of differential end test mass motion. The interferometer senses and digitizes a change in DARM length, $\Delta L_{ext}$ according to $\gamma(t)C_{L}(f)$, the result of which is the digital error signal $e_{D}$, which is then fed back through a set of digital filters $D(f)$, and converted to analog control via the actuation function of the end test masses $A(f)$ . \label{darmloop}}
\end{figure}

Feedback control systems are used to hold the interferometer in a regime where the digital error signal, $e_{D}(f)$, is linearly related to the DARM length, $\Delta L_{ext}$, (as in Eq. \ref{lineardarmloop}) and hence to the gravitational wave strain, $h(f)$. We model this control loop as a single-input, single-output control loop depicted in Figure \ref{darmloop}. 

The loop contains three major subsystems. First is the length sensing function, $C_{L}(f,t)$, which describes how the interferometer responds to differential changes in arm lengths and how that response is digitized. This function is separated into a frequency-dependent function $C_{L}(f)$ which may have some slow time dependence captured by a factor $\gamma(t)$. $D(f)$ is a set of digital filters, used to shape the loop error signal into a control signal. The remaining subsystem  is the actuation function, $A(f)$, which describes how the test masses physically respond to the digital control signal.  We assume linear relationships between all subsystems, such that any subsystem (and internal components) may be defined by the ratio of output over input signals. 

The product of frequency-dependent subsystems inside the control loop is the ``open loop transfer function'' $G_{L}(f)$,
\begin{eqnarray}
G_{L}(f) & = &  C_{L}(f) D(f) A(f). \label{olgdef}
\end{eqnarray}
Using the above model, we derive the length response function, $R_{L}(f,t)$, in terms of these functions to be
\begin{eqnarray}
R_{L}(f,t) & \equiv &  \frac{1 + \gamma (t) G_{L}(f)}{\gamma (t) C_{L}(f)}. \label{RL}
 \end{eqnarray}
The remainder of this section describes the components of each subsystem in the control loop.

\subsection{Sensing Function}
\label{sen}
\begin{figure}[h!]
\begin{center}
\includegraphics[width=90mm]{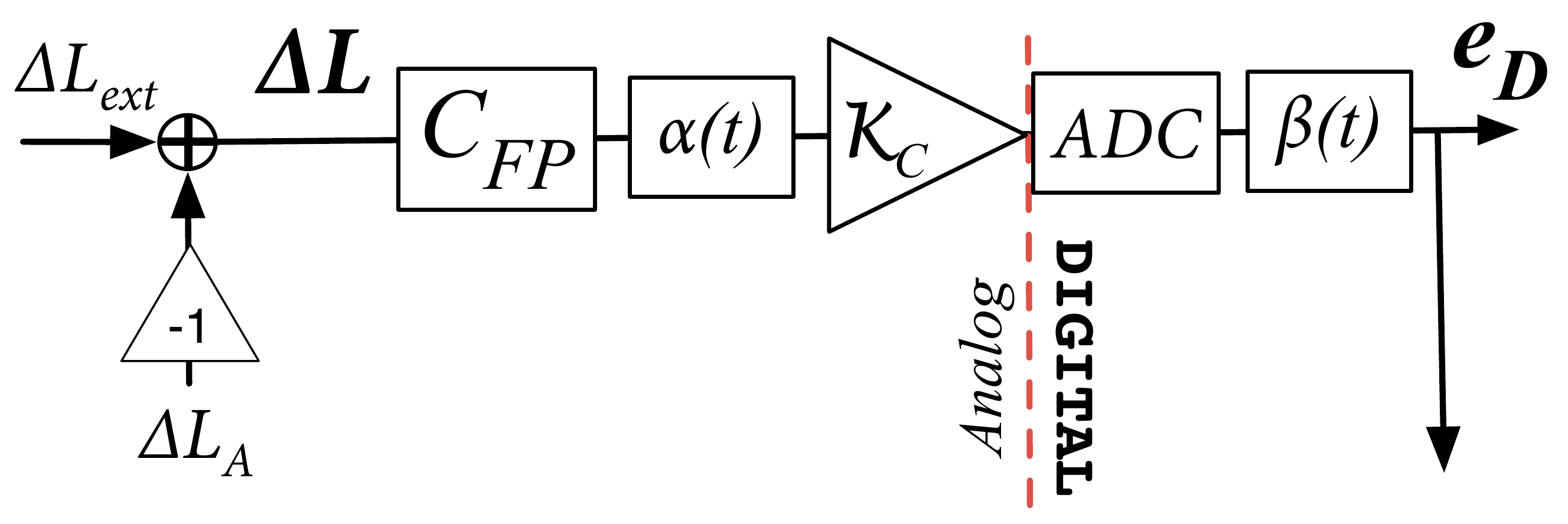}
\end{center}
\caption{Schematic breakdown of the sensing function $C_{L}(f,t)$. Internal to the LIGO Scientific Collaboration, the digital signal $e_{D}$ is often colloquially referred to by its digital ``channel'' name DARM\_ERR. From left to right, $C_{FP}(f) \propto H_{FP}^{x} + H_{FP}^{y}$ is the arm cavity transfer function; $\alpha(t)$ is the time-dependent variation of the interferometer's input laser power and optical gain; $\mathcal{K}_{C}$ is the scaling coefficient which absorbs all constants including the input laser power, optical gain, the quantum efficiency of the photodiodes, the impedance of the photodiode circuitry, and the analog-to-digital gain; $ADC(f)$ is the frequency dependence of the analog to digital conversion; and $\beta(t)$ is the digital factor which compensates for the analog change $\alpha(t)$. The compensation is not perfect, therefore the factor $\gamma(t) \equiv \alpha(t)\beta(t)$ represents the residual variation.  \label{sensdetail}}
\end{figure}

The length sensing function, $C_{L}(f,t)$, describes the transfer function between the residual change in DARM length, $\Delta L(f)$, and the digital error signal, $e_{D}(f)$,
\begin{eqnarray}
C_{L}(f,t) & = & \gamma(t) \frac{e_{D}(f)}{\Delta L(f)}. \label{CLdef}
\end{eqnarray}
It is important to note that this linear relationship between the DARM length change and the digital error signal only applies when the detector is under control of the feedback loop: in Eq.~\ref{CLdef}, $\Delta L(f)$ is the residual external DARM length change, $\Delta L_{ext}(f)$, after the controlled length change, $\Delta L_{A}(f)$, is applied. The sensing function has several components  (shown in Figure \ref{sensdetail}) which are treated independently,
\begin{eqnarray}
C_{L}(f,t) & = & \gamma(t) \times \mathcal{K}_{C} \times \bigg[C_{FP}(f) ~\times~ ADC(f)  \bigg]. \label{sensdef}
\end{eqnarray}
The constant, $\mathcal{K}_{C}$, which holds all frequency-independent scaling factors, has dimensions of digital counts of error signal per unit change in DARM length. The remaining terms in Eq. \ref{sensdef} are dimensionless, including time dependence, treated independently in the coefficient $\gamma(t)$. 

The change in each arm cavity length $L$ affects the phase of the laser's electric field returning from the cavity. On resonance, the transfer function between the change in electric field phase reflected by the cavity input mirror $\Phi(f)$ and a change in cavity length is
\begin{eqnarray}
H_{FP}(f) & = &  \frac{2 \pi}{\lambda} \frac{1}{r_{c}} \frac{r_{e}(1 - r_{i}^2)}{(1-r_{i}r_{e})}\frac{\sin(2 \pi f L / c)}{2 \pi f L /c}\frac{e^{-2 \pi i f L /c}}{1 - r_{i}r_{e}e^{-4 \pi i f L / c}} \label{FP}
\end{eqnarray}
where $\lambda$ is the laser wavelength, $r_{c} = (r_{e} - r_{i})/(1-r_{i}r_{e})$ is the on-resonance Fabry-Perot arm cavity reflectivity, $r_{i}$ and $r_{e}$ are the amplitude reflectivity of the input and end test masses, and $c$ is the speed of light. In the frequency band considered for analysis, where 40 Hz $< f <$ 6 kHz $ \ll 2c/L$, the frequency-dependence of $H_{FP}(f)$ is approximated by a simple ``cavity pole'' transfer function,
\begin{eqnarray}
H_{SP}(f) & \equiv & \frac{H_{FP}(f \ll c/2L)}{ H_{FP}(0)} ~\approx~ \frac{1}{1+i \frac{f}{f_{c}}}, \label{SP}
\end{eqnarray}
where $f_{c} = c~(1 - r_{i}r_{e}) / 4 \pi L \sqrt{r_{i} r_{e}}$ \cite{pdh3, hfcorrs, Sensing2, peanut,freqresp}.

The LIGO detectors use a Pound-Drever-Hall detection scheme to extract this phase information from the arm cavities, which is recombined at the beam splitter. The laser electric field input into the interferometer is phase-modulated at $\omega_{m}/2\pi = $25 MHz, which effectively splits the field into a ``carrier'' field with the original laser frequency, $\Omega$, and upper and lower ``sideband'' fields with frequency $\Omega \pm \omega_{m}$. The sideband fields resonate in the power recycling cavity but are anti-resonant in the arm cavities, and therefore, unlike the carrier field, experience no phase change from the arm cavity length variation. The Michelson is set up with a fixed asymmetry such that, at the anti-symmetric port, the carrier field is held on a dark fringe and the sideband fields are not. In this setup, when the arm cavity lengths change differentially, the carrier field moves away from the dark fringe, mixes with the sideband field at the antisymmetric port, and a beat signal at $\omega_{m}$ is generated.

The power of the mixed field at the antisymmetric port (in Watts) is sensed by four photodiodes. The photocurrent from these diodes is converted to voltage, and then demodulated at 25 MHz. This voltage signal (and therefore the change in DARM length) is proportional to power of the input laser field, the ``optical gain'' (the product of Bessel functions of modulation strength, the recycling cavity gain,  the transmission of the sidebands into the antisymmetric port from the Michelson asymmetry, the reflectivity of the arm cavities for the carrier), the quantum efficiency of the photodiodes, and the impedance of the photodiode circuitry \cite{pdh2,pdh3,peanut,freqresp}. The demodulated voltage from the photodiodes is whitened, and anti-aliased with analog circuitry and then digitized by an analog-to-digital converter which scales the voltage to digital counts. The frequency dependence of the anti-aliasing filters and digitization process is folded into the function $ADC(f)$. We absorb all proportionality and dimensions of this process into the single constant, $\mathcal{K}_{C}$, having dimensions of digital counts per meter of DARM test mass motion. 

The optical gain is time-dependent because small, low-frequency ($f \ll $ 40 Hz) alignment and thermal lensing fluctuations in the resonant cavities change the carrier and sideband field amplitudes. The input laser power may also fluctuate from similar alignment and thermal effects. We represent these variations with a coefficient, $\alpha(t)$. The input power, along with the carrier and sideband power stored in the cavities, are monitored by several independent photodiodes. Their signals are also digitized and combined to form a coefficient, $\beta(t)$, used to digitally compensate for the time-dependent variations. The compensated anti-symmetric port signal forms the error signal for the DARM control loop, $e_{D}(f)$. The sensing function therefore depends on both time and frequency, but can be separated into independent components $C_{L}(f,t) = \gamma(t) C_{L}(f)$, where  
\begin{eqnarray}
C_{L}(f) & \propto & C_{FP}(f) \times ADC(f) = \left[H_{SP}^{x}(f) + H_{SP}^{y}(f)\right] \times ADC(f) \label{CL}
\end{eqnarray}
and 
\begin{eqnarray}
\gamma(t) \equiv \alpha(t) \beta(t) 
\end{eqnarray}
is the scale factor of order unity accounting for the residual time dependence after compensation.

\subsection{Digital Filters}
\label{dig}

The digital filters, $D(f)$, are known functions in the model. These filters are used to shape the digital DARM control loop error signal, $e_{D}(f)$ (in digital counts proportional to displacement) into a digital control signal, $s_{D}(f)$ (in digital counts proportional to force),
\begin{eqnarray}
D(f) & = & \frac{s_{D}(f)}{e_{D}(f)}
\end{eqnarray}

 Over the course of the science run, discrete changes are made to the digital filters, $D(f)$, to improve the performance and stability of the detector (four times in the Hanford interferometers, three in Livingston). These changes significantly alter the frequency-dependence of the DARM control loop, and hence affect the overall response function of the interferometer. We divide the run into ``epochs'' defined by these changes. 

Note that the digital filter component does not include all digital filters in the DARM loop. Both the sensing function and the actuation function contain digital filters, but their frequency dependence is either negligible in the measurement band, only important in a very narrow frequency range, or are compensating for analog circuitry whose product with the digital filters form a unity transfer function. We include these filters in their respective sub-systems for completeness.

\subsection{Actuation function}
\label{act}
\begin{figure}[h!]
\begin{center}
\includegraphics[width=90mm]{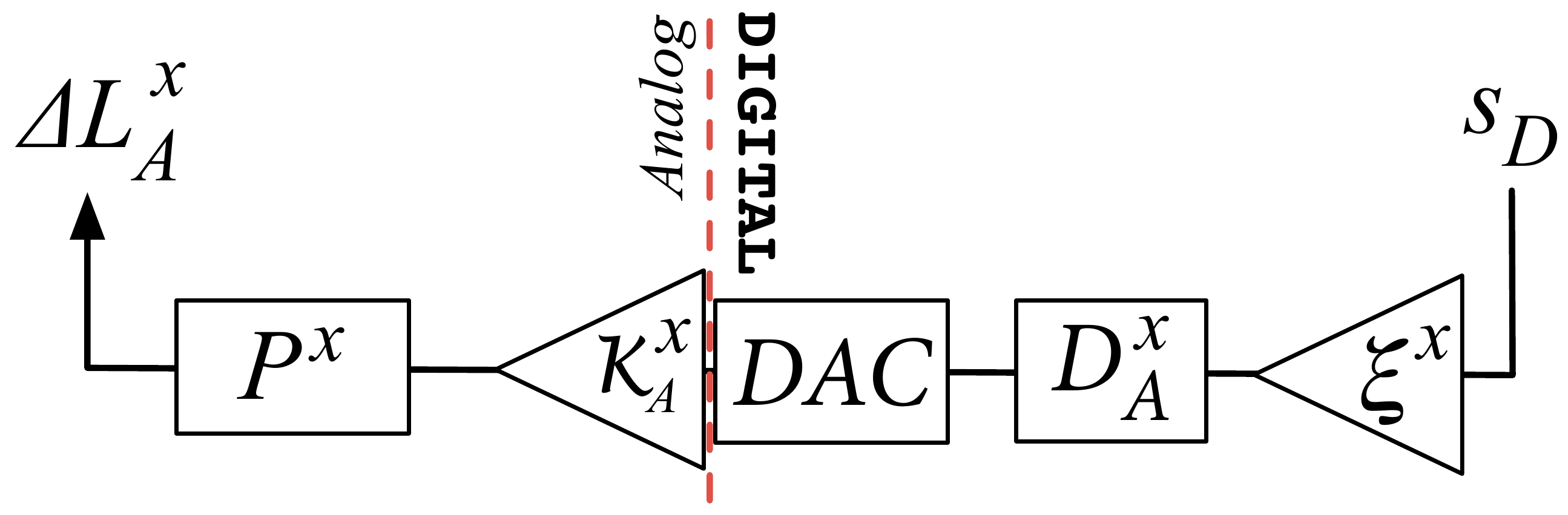}
\end{center}
\caption{Schematic breakdown of the signal flow through the actuation function for the X arm $A^{x}(f)$. The digital signal $s_{D}(f)$ is colloquially referred to by its ``channel'' name DARM\_CTRL. From right to left, $\xi^{x}$ is the fraction of the digital control signal sent to the X arm; $D_{A}^{x}(f)$ are digital filters; $DAC(f)$ is the frequency dependence of digital to analog conversion;  $K_{A}^{x}$ is the scaling coefficient proportional to the digital-to-analog gain, the gain of resistance circuitry which converts voltage to current, the gain of the coil actuators which convert current to magnetic force, and the force-to-displacement transfer function gain; and $P^{X}$ is the frequency dependence of the force-to-displacement transfer function.  \label{actdetail}}
\end{figure}

The actuation function $A(f)$ is defined by the transfer function between the digital control signal, $s_{D}(f)$, and the physical motion imposed on the end test masses by the control loop, $\Delta L_{A}$,
\begin{eqnarray}
A(f) & = & \frac{ \Delta L _{A}}{s_{D}(f)},
\end{eqnarray}
and has units of end test mass displacement in meters per count of digital control signal.  We describe the actuation function as a linear combination of functions for each test mass, 
\begin{eqnarray}
A(f) = \xi^{x} A^{x}(f) + \xi^{y} A^{y}(f). \label{actsplitdef}
\end{eqnarray}
where $\xi^{x,y}$ are known digital coefficients of order unity, roughly equivalent, but opposite in sign. Once split, the control signal flows through each component to the end test masses as shown in Figure \ref{actdetail}.

For each arm, the digitally split control signal passes through digital suspension filters, $D_{A}(f)$, and is converted from digital signal to an analog voltage via the digital to analog conversion element, $DAC(f)$, which includes analog anti-imaging circuitry. The resulting voltage passes through a resistance circuit converting it into current, and is sent to the coil actuators which convert the current into force on the magnets attached to the end test mass. The suspended test mass is displaced according to the force-to-displacement transfer function, $P(f)$, changing each arm cavity length, $\Delta L_{A}(f)$. The arm's scaling coefficients, $\mathcal{K}_{A}$, absorb all dimensions and frequency-independent factors in the actuation path. This includes the digital-to-analog gain, the gain of the resistance circuitry, the gain of the coil actuators, and the force-to-displacement transfer function scale factor. In summary, we express the individual end test mass actuation functions in Eq. \ref{actsplitdef} as
\begin{eqnarray}
A^{x,y}(f) & = & \mathcal{K}_{A}^{x,y} \times \Big[D_{A}^{x,y}(f) \times DAC(f) \times P^{x,y}(f)  \Big]. \label{aef}
\end{eqnarray}
The actuation coefficients, $\mathcal{K}_{A}$, scale the arbitrary counts of digital excitation force into meters of test mass motion. The remaining terms in Eq. \ref{aef} are dimensionless.

\begin{figure}[h!]
\begin{center}
\includegraphics[width=60mm]{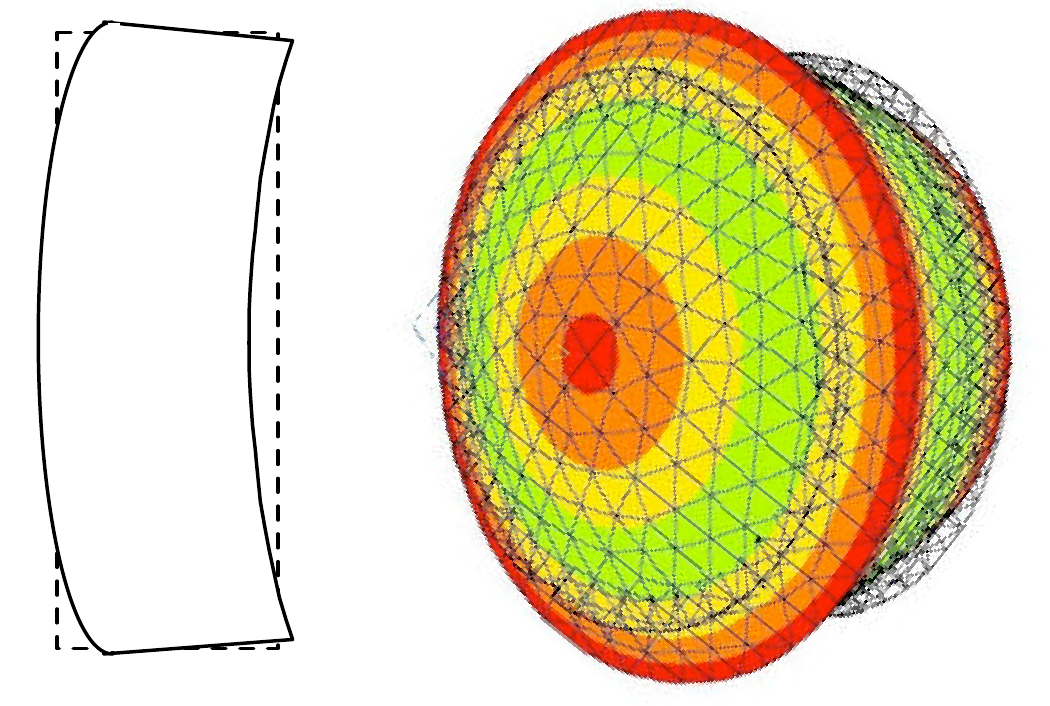}
\end{center}
\caption{Physical shape of the end test mass drumhead internal resonance. Left: Cartoon, edge-on view of the fundamental mode of a cylindrical plate \cite{acoustics}. Right: Three dimensional modal shape of the drumhead resonance from finite element analysis of a cylinder with dimensions similar to the LIGO test masses \cite{tmmodes}.  \label{dhm}}
\end{figure}

The suspended test mass can be treated as a pendulum driven by the coil actuators (see Figure \ref{sus}). The force-to-displacement transfer function for the center of mass of a pendulum, $P_{cm}(f)$, is 
\begin{eqnarray}
P_{cm}(f) & \propto & \frac{1}{[f_{0}^{cm}]^{2} + i \frac{[f_{0}^{cm}]}{Q^{cm}}~f - f^{2}} \label{pendtf}
\end{eqnarray}
where $f_{0}^{cm}$ and $Q^{cm}$ are the frequency and quality factor of the pendulum. A rigid body resonant mode akin to the fundamental mode of a cylindrical plate \cite{acoustics} (see Figure \ref{dhm}) known as the ``drumhead'' mode is also included in the force-to-displacement model. Its radially-symmetric shape, excited by the actuators, lies directly in the optical path and amplifies the cavity's response to the length control signal above a several kHz \cite{tmmodes}.  We approximate the effects of the resonance by multiplying $P_{cm}(f)$ by an additional pendulum transfer function,$P_{dh}(f)$, defined by frequency, $f_{0}^{dh}$, and quality factor, $Q^{dh}$. The total force-to-displacement transfer function is
\begin{eqnarray}
P(f) & \propto & P_{cm}(f) P_{dh}(f)
\end{eqnarray}

The digital suspension filters, $D_{A}(f)$, are between the split control signal and the digital-to-analog converter. Their purpose is to remove control signal in narrow frequency ranges around the frequencies of other in-band, non-axisymmetric, rigid-body resonant modes of the test masses that are excited by the actuation forces \cite{tmmodes}, and to reduce the coupling between DARM length motion and angular motion of the test mass.

\section{Measurements}
\label{meas}
Each subsystem of the response function $R_{L}(f)$ is developed using measurements of key parameters in their modeled frequency dependence and their scaling coefficients. The digital filter subsystem is completely known; its frequency dependence and scaling coefficient are simply folded into the model of the response function. The parameters of the frequency-dependent portions of the sensing and actuation subsystems may be obtained precisely by direct measurement or are known from digital quantities and/or design schematics. As such, these parameters' measurements will only be briefly discussed. 

The detector's sensing function behaves in a non-linear fashion when uncontrolled, therefore we may only infer the linear model's scaling coefficient, $\mathcal{K}_{C}$, from measurements of the detectors under closed control loops. We infer that the remaining magnitude ratio between our model and measurements of the open loop transfer function $G_{L}(f_{UGF})$ as the sensing coefficient $\mathcal{K}_{C}$ (where $f_{UGF}$ is the unity gain frequency of the DARM control loop). Other than the known frequency-independent magnitude of $D(f)$, the open loop gain model's magnitude is set by the actuation scale factor, $\mathcal{K}_{A}$. This makes it a crucial measurement in our model because it sets the frequency-independent magnitude of the entire response function. Measurements of the open loop transfer function over the entire gravitational wave frequency band are used to confirm that we have modeled the correct frequency dependence of all subsystems. Finally, measurements of $\gamma(t)$ track the time dependence of the response function. The details of these measurements and respective uncertainty estimates are described below.

\subsection{Actuation Function}
\label{actmeas}
The components of each arm's actuation function, $\mathcal{K}_{A}^{x,y}$, $DAC(f)$ and $P^{x,y}(f)$ are measured independently in a given detector. As with $D(f)$, both $\xi^{x,y}$ and $D_{A}^{x,y}(f)$ are digital functions included in the model without uncertainty.

\subsubsection{Actuation Scaling Coefficients, $\mathcal{K}_{A}^{x,y}$}
\label{dccal}
The standard method for determining the actuation coefficients, $\mathcal{K}_{A}^{x,y}$, used for the fifth science run is an interferometric method known as the ``free-swinging Michelson'' technique; a culmination of several measurements with the interferometer in non-standard configurations. The method uses the interferometer's well-known Nd:YaG laser wavelength ($\lambda = 1064.1 \pm 0.1$ nm, \cite{lambda1,lambda2}) as the calibrated length reference while using the test mass' coil actuators to cause a length change. Details of the technique are described in \ref{simpmichappendix}. The actuation coefficient is measured using this method many times for each optic in each interferometer over the course of the science run, and their mean used as the actuation scaling coefficient for all model epochs. Table \ref{dccals} summarizes the actuation coefficients, $\mathcal{K}_{A}^{x,y}$, for the three interferometers in the fifth science run, using free-swinging techniques. 

\begin{table}[h!]
\caption{Summary of the actuation scaling coefficients measured during S5. These single numbers are formed by the mean of each measurement's median $\langle \mathcal{K}_{A}^{x,y} \rangle_{j}$ (6 for each end test mass in H1, 5 in H2, and 14 and 15 for the X and Y test masses, respectively in L1). Only statistical uncertainty is reported here; systematic uncertainty is folded the total uncertainty of the actuation function. \label{dccals}}
\begin{center}
\begin{tabular}{|c|c|c|}
\hline
 & $  \mathcal{K}_{A}^{x} $ (nm/ct) & $  \mathcal{K}_{A}^{y} $ (nm/ct) \\
\hline
H1 & $0.847 \pm 0.024$ & $0.871 \pm 0.019$\\
H2 & $0.934 \pm 0.022$ & $0.958 \pm 0.034$\\
L1 & $0.433 \pm 0.039$ & $0.415 \pm 0.034$\\
\hline
\end{tabular}
\end{center}
\end{table}

\subsubsection{Force-to-Displacement Transfer Function, $P(f)$}

Each test mass coil actuator system is equipped with an optical position sensor system that consists of an infrared LED emitter aimed at a small photodiode mounted in the coil actuator, and a mechanical ``flag'' attached to the magnet on the optic that cuts through the beam. From amplitude spectral densities of these sensor signals while the optic is free-swinging, the frequency of each center-of-mass transfer function, $f_{0}^{cm}$ is measured with negligible uncertainty. The quality factors, $Q^{cm}$, depend on the amount of local damping applied to suspension, but are estimated from driven transfer functions. The uncertainty of this estimation, though large, has little effect on the center-of-mass transfer function in the frequency band of interest and is ignored. Table \ref{fsandqs} shows the results for the center of mass force-to-displacement transfer function. The drumhead frequency, $f_{0}^{dh}$ for each test mass in the Hanford and Livingston detectors have been measured to be 9.20 kHz and 9.26 kHz respectively with $Q^{dh}\sim10^{5}$ \cite{tmmodes,tmres}, where again though the uncertainty in these parameters may be large, it has little effect in band and is ignored.

\begin{table}[h!]
\caption{Summary of pendulum frequencies, $f_{0}^{cm}$, and quality factors, $Q^{cm}$, used to compose models of each interferometer's center-of-mass pendulum transfer functions in S5.\label{fsandqs}}
\begin{center}
\begin{tabular}{|c|c|c|c|c|}
\hline
\multirow{2}{*}{{\bf }} & \multicolumn{2}{|c|}{{\bf X End Test Mass}} & \multicolumn{2}{c|}{{\bf Y End Test Mass}}\\ 
\hline
 & $f_{0}^{cm}$ (Hz) & $Q^{cm}$ &  $f_{0}^{cm} $(Hz) & $Q^{cm}$ \\ \hline
H1 & 0.767 & 10 & 0.761 & 10 \\
H2 & 0.749 & 10 & 0.764 & 10 \\
L1 & 0.766 & 100 & 0.756 & 100 \\
\hline
\end{tabular}
\end{center}
\end{table}

\subsubsection{Digital to Analog Conversion, $DAC(f)$}
The digital to analog conversion model $DAC(f)$, includes the effects of the finite sample-and-hold method used to convert digital signal to an analog voltage, the analog anti-imaging filter, measured residual frequency dependence from imperfect digital compensation of analog de-whitening, and the time delay arising from computation and signal travel time. 

We use the standard model for the sample and hold of the digital to analog converter \cite{oppnschaf,christi}
\begin{eqnarray}
H_{s}(f) & = & \textrm{sinc}\Big[(2 \pi f) / (2 f_{s})\Big]~e^{- i (2 \pi f) / (2 f_{s})},
\end{eqnarray}
where the sample frequency  $f_{s} = 16384$ Hz is used in all detectors. 

The same analog anti-image filter is used for each of the four coils on the test mass. They are analog, third-order, Chebyshev low-pass filters with 0.5 dB passband ripple whose corner frequency is at 7.5 kHz and 8.1 kHz for the Hanford and Livingston detectors, respectively, and modeled as such in the $DAC(f)$ transfer function. We also include residuals measured between the modeled anti-imaging filter and its analog counterpart.

For a given end test mass, there is a complementary pair of digital and analog whitening filters for each of the four coil actuators. A comparison between the digital compensation and the real analog electronics has shown non-negligible, frequency-dependent residuals. We measure the residuals for all four coils in each test mass by taking the ratio of transfer functions between a digital excitation and the analog output of the whitening filters with the digital filters on and off. We include the average residual of the four coils in our model. 

A detailed analysis of the digital time delay in the digital-to-analog conversion has been performed elsewhere \cite{timing}. For the actuation model we estimate the time delay from our model of the open loop transfer function (attributing all residual delay in the loop to the actuation function), and assign a fixed delay to each epoch. 

\subsubsection{Actuation Uncertainty, $\sigma_{A}$}
\label{acterror}
The digital suspension filters, $D_{A}(f)$, have well-known digital transfer functions, which are included in the model without an uncertainty. The model of force-to-displacement transfer function, $P(f)$, and digital-to-analog conversion, $DAC(f)$, are derived from quantities with negligible uncertainty. Hence, the uncertainty estimate for the actuation function is derived entirely from measurements of the actuation scaling coefficient, $\mathcal{K}_{A}$.

The actuation coefficient is measured using a series of complex transfer functions taken to be frequency independent as described in \ref{simpmichappendix}. We take advantage of this fact by estimating the frequency-independent uncertainty in the overall actuation function from the statistical uncertainty of all free-swinging Michelson measurements. For magnitude, we include a systematic uncertainty originating from an incomplete model of the actuation frequency-dependence, such that the total actuation uncertainty is
\begin{eqnarray}
\left(\frac{\sigma_{|A|}}{|A|}\right)^{2} & = & \left(\frac{\sigma_{|\mathcal{K}_{A}|}}{|\mathcal{K}_{A}|}\right)^{2} + \left(\frac{\sigma_{(r/a)}}{(r/a)}\right)^{2}  \\
\sigma_{\phi_{A}}^{2} & = & \sigma_{\phi_{\mathcal{K}_{A}}}^{2}. \label{actunc}
\end{eqnarray}

The statistical uncertainties, $\sigma_{|\mathcal{K}_{A}|}/|\mathcal{K}_{A}|$ and $\sigma_{\phi_{\mathcal{K}_{A}}}$, are the quadrature sum of the scaling coefficient uncertainty from each test mass, as measured by the free-swinging Michelson technique. For each optic's coefficient, we estimate the uncertainty by taking the larger value of either the standard deviation of all measurement medians, or the mean of all measurement uncertainties divided by the square root of the number of frequency points in a given measurement. These two numbers should be roughly the same if the measured quantity followed a Gaussian distribution around some real mean value and stationary in time. For all optics, in all interferometers, in both magnitude and phase, these two quantities are not similar, implying that the measurements do not arise from a parent Gaussian distribution. We attribute this to the quantity changing over time, or a systematic error in our measurement technique that varies with time.  Later studies of the free-swinging Michelson technique have revealed that the probable source of this time variation is our assumption that the optical gain of the simple Michelson remains constant over the measurement suite (see \ref{simpmichappendix}). 

We have folded in an additional $\sigma_{(r/a)} / (r/a) = $ 4\% systematic error in magnitude for the Hanford detectors only. This correction results from the following systematic difference between the Hanford and Livingston free-swinging Michelson measurement setup. Analog suspension filters, common to all detectors, are used to increase the dynamic range of the coil actuators during initial control of the test masses.  When optic motions are sufficiently small enough to keep the cavity arms on resonance, they are turned off and left off as the detectors approach designed sensitivity \cite{detectorpaper,locking}. These additional suspension filters were left in place for the Hanford measurements in order to obtain better signal-to-noise ratios for the driven transfer functions described in \ref{simpmichappendix}. The filters' color had been compensated with digital filters, but the average residual frequency dependence is roughly 4\% for both end test masses in H1 and H2.

\begin{figure}[h!]
\begin{center}
\includegraphics[width=90mm]{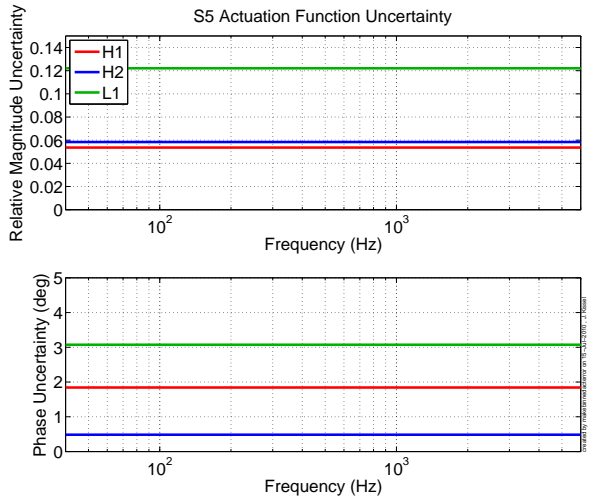}
\end{center}
\caption{Summary of the actuation uncertainty for all detectors in S5.  \label{totalactunc}}
\end{figure}

The total uncertainty for each interferometer's actuation function, as described in Eq. \ref{actunc}, is shown in Figure \ref{totalactunc}. These estimates include statistical and known systematic uncertainties. To investigate potential unknown systematic uncertainties in the actuation functions we applied two fundamentally different calibration methods. The results of these investigations are described in section 5.

\subsection{Sensing Function}
\label{sensmeasts}
The components of the sensing function, $\mathcal{K}_{C}$, $C_{FP}(f)$, and $ADC(f)$ are described in \S \ref{sen}. The frequency-dependent components are developed from measured parameters with negligible uncertainties, and $K_{C}$ is obtained as described above. The techniques used to obtain the parameters are described below.

\subsubsection{Sensing Scaling Coefficient, $\mathcal{K}_{C}$}
In principle, the scaling coefficient $\mathcal{K}_{C}$ is also composed of many independently measurable parameters as described in \S \ref{sen}. In practice, these components (specifically components of the optical gain) are difficult to measure independently as the interferometer must be controlled into the linear regime before precise measurements can be made. The scaling coefficient for the other subsystems are either measured (in the actuation) or known (in the digital filters). We take advantage of this by developing the remainder of sensing subsystem (i.e. its frequency-dependence), forming the frequency-dependent loop model scaled by the measured actuation and known digital filter gain, and assume the remaining gain difference between a measurement of open loop transfer function and the model is entirely the sensing scale factor. Results will be discussed in \S\ref{olg}.

\subsubsection{Fabry-Perot Cavity Response, $C_{FP}(f)$}
Our model of the Fabry-Perot Michelson frequency response is the sum of the response from each arm as in Eq. \ref{CL}. Using the single pole approximation (Eq. \ref{SP}), the frequency response of each arm cavity $H_{SP}^{x,y}(f)$ can be calculated explicitly using a single measured quantity, the cavity pole frequency $f_{c}$. We compute $f_{c}$ by measuring the light storage time $\tau = 1/(4 \pi f_{c})$ in each cavity. 

A single measurement of the storage time is performed by aligning a single arm of the interferometer (as in the right panel of Figure \ref{simplemich}) and holding the cavity on resonance using the coil actuators. Then, the power transmitted through that arm is recorded as we rapidly take the cavity out of resonance. We fit the resulting time series to a simple exponential decay, whose time constant is the light storage time in the cavity. This measurement is performed several times per arm, and the average light storage time is used to calculate the cavity pole frequency. Table \ref{fctable} shows the values of $f_{c}$ used in each model.

\begin{table}[h!]
\caption{Summary of cavity pole frequencies $f_{c}$ used in each interferometer's sensing function in S5. H1 and H2 have used the average of each arm, hence their numbers reported below are the same, with uncertainty estimated as the quadrature sum of each result.\label{fctable}}
\begin{center}
\begin{tabular}{|c|c|c|}
\hline
 & $f_{c}^{x}$ (Hz) & $f_{c}^{y}$ (Hz) \\
\hline
H1 & 85.6 $\pm$ 1.5 & 85.6 $\pm$ 1.5 \\
H2 & 158.5 $\pm$ 2.0 & 158.5 $\pm$ 2.0 \\
L1 & 85.1 $\pm$ 0.8   & 82.3 $\pm$ 0.5 Hz \\
\hline
\end{tabular}
\end{center}
\end{table}

\subsubsection{Analog to Digital Conversion, $ADC(f)$}
Each of the four photodiodes used to measure the power at the dark port are sampled at 16384 Hz. The dominant frequency dependence of this analog-to-digital conversion process arises from the analog anti-aliasing filters. These filters are analog eighth order elliptic filters, which differ only in corner frequency at the two sites: 7.5 kHz for the Hanford and 8.1 kHz for Livingston. The frequency dependence is unity below 1 kHz. Above a few kHz, the magnitude changes less than 2\%, but the phase loss from these filters becomes non-negligible ($>$ 180 deg). The residual frequency dependence between this model and measured transfer function of the filter is also included. The discrepancy occurs only above 1 kHz and varies less than 2\% in magnitude and 5 degrees in phase.

\subsubsection{Time Dependence, $\gamma(t)$}
\label{gamma}
\begin{figure}[h!]
\begin{center}
\includegraphics[width=90mm]{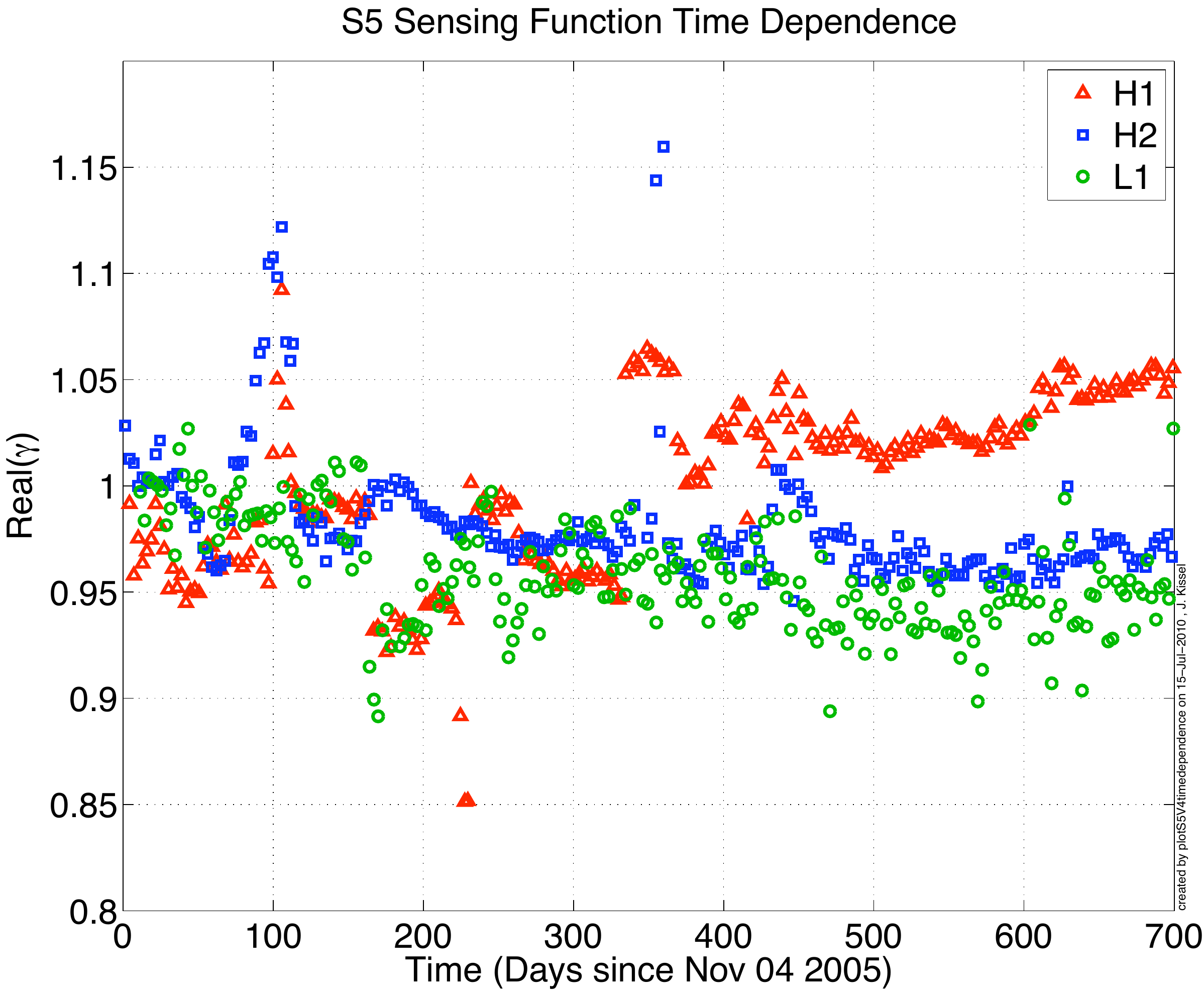}
\end{center}
\caption{Time-dependent corrections to the sensing function $\Re e\{\gamma(t)\}$ over the course of the science run. \label{regamma}}
\end{figure}

We measure the time dependence of the sensing function by digitally injecting a signal, $s_{cl}(f)$, at the output of the digital filters, $D(f)$, prior to the control signal, $s_{D}(f)$, at three line frequencies $f_{cl}$ near 50, 400, and 1100 Hz. The time-dependent coefficient $\gamma(t)$ is defined as
\begin{eqnarray}
\gamma(t) & = & \alpha(t)\beta(t) ~=~ - \frac{1}{G_{L}(f_{cl})} \frac{s_{D}(f_{cl}) - s_{cl}(f_{cl})}{s_{D}(f_{cl})} , \label{gammadef}
\end{eqnarray}
where $G_{L}(f_{cl})$ is the modeled DARM open loop transfer function at the reference time in each epoch at a given calibration line frequency, $f_{cl}$; $s_{cl}(f_{cl})$ and $s_{D}(f_{cl})$ are the excitation signal and the control signal, respectively, each digitally demodulated at the same frequency and averaged over 60 seconds.  The coefficient generated from $f_{cl} \approx$ 400 Hz is used to scale the response function model; the other two frequencies are used to confirm that the variations are independent of frequency.  In the ideal case (no noise on top of the injected line and with a perfect model for $G_{L}(f_{cl})$), the coefficient is a real factor near unity. Figure \ref{regamma} shows the evolution of $\Re e\{\gamma(t)\}$ over the course of the science run for each detector. 
 
We also separate the relative uncertainty of the time dependent coefficient $(\sigma_{\gamma} / \gamma)^{2}$ into those of systematic and statistical origin. As the coefficient is ideally real and unity, we expect the imaginary part of the measurement defined in Eq. (\ref{gammadef}) to be a random time series with zero mean. A non-zero mean would indicate a systematic error in our estimate of $\Re\{\gamma(t)\}$, given by $\sqrt{1 - \left(\overline{\Im m\{\gamma(t)\}}\right) ^{2}}$ (assuming the real part of $\gamma(t)$ is unity). The measured mean is less than 5\% for all detectors, implying a negligible systematic error of 0.1\% and is ignored.

\begin{figure}[h!]
\begin{center}
\includegraphics[width=90mm]{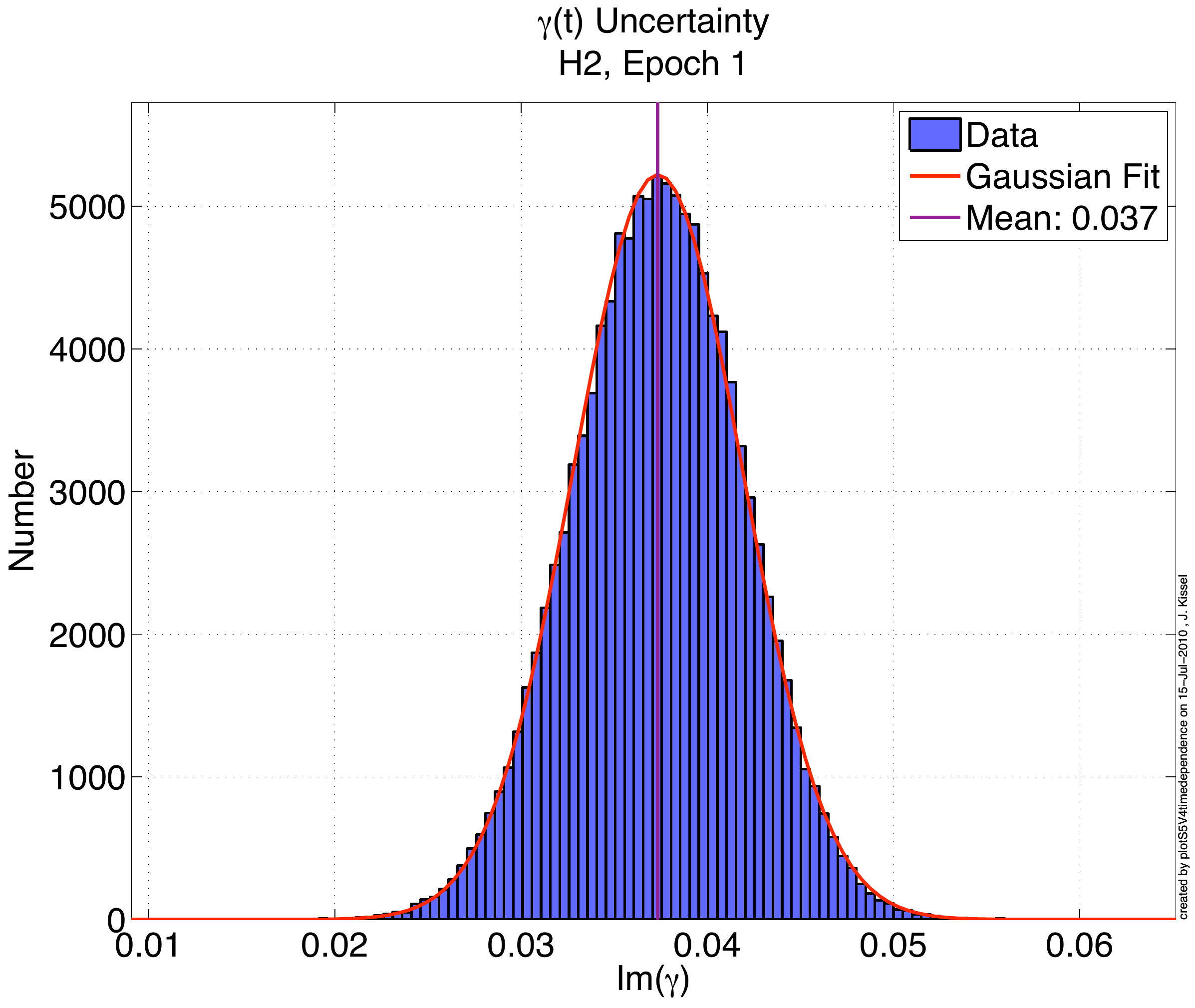}
  \caption{Histogram of $\Im m\{\gamma(t)\}$ for the epoch 1 in H2, the standard deviation of which represents the error estimation $\sigma_{\gamma}$ for this epoch. \label{imgamma}}
  \end{center}
  \end{figure}

The statistical error is determined by the signal-to-noise ratio of the calibration line at frequency $f_{cl}$, and is estimated by the standard deviation of $\Im m\{\gamma(t)\}$, measured in every epoch at a sampling rate of 1 Hz.  Though the statistical error is roughly equivalent in all epochs for a given detector, we chose the largest standard deviation as a representative error for the entire run. Figure \ref{imgamma} shows an example histogram of $\Im m\{\gamma(t)\}$ for H2.
 
\subsection{Open Loop Transfer Function}
\label{olg}
The open loop transfer function, $G_{L}(f)$, is measured while the interferometer is controlled, operating in the nominal configuration, at designed sensitivity. We use a digital DARM excitation with amplitude much larger than  $\Delta L_{ext}$, such that we may assume it to be a contribution to measurement noise. During the measurement we assume no time-dependent variations occur, and set $\gamma(t) = 1$. We compare this measurement against our model of the open loop transfer function which is the product of each subsystem described above (see Eq. \ref{olgdef}), and scale the model by the measurement's magnitude at the expected unity gain frequency to form $\mathcal{K}_{C}$ as described in \S \ref{sensmeasts}. Values for the sensing scaling coefficient averaged over epochs, are shown in Table \ref{Cc}.

\begin{table}[h!]
\caption{Average value for scaling coefficients $\mathcal{K}_{C}$ for the sensing function, $C_{L}$(f,t) for each interferometer. They are stated without uncertainty, since these quantities are derived from measurements of the open loop gain and actuation scaling coefficient. See further discussion in \S\ref{error}.  \label{Cc}}
\begin{center}
\begin{tabular}{|c|c|}
\hline
 & $\mathcal{K}_{C}~(cts/10^{-15} m)$\\
\hline
H1 & $0.15$ \\
H2  &  $0.61$ \\
L1 &  $9.1$ \\
\hline
\end{tabular}
\end{center}
\end{table}

\begin{figure}[h!]
  \begin{minipage}{70mm}
\includegraphics[width=70mm]{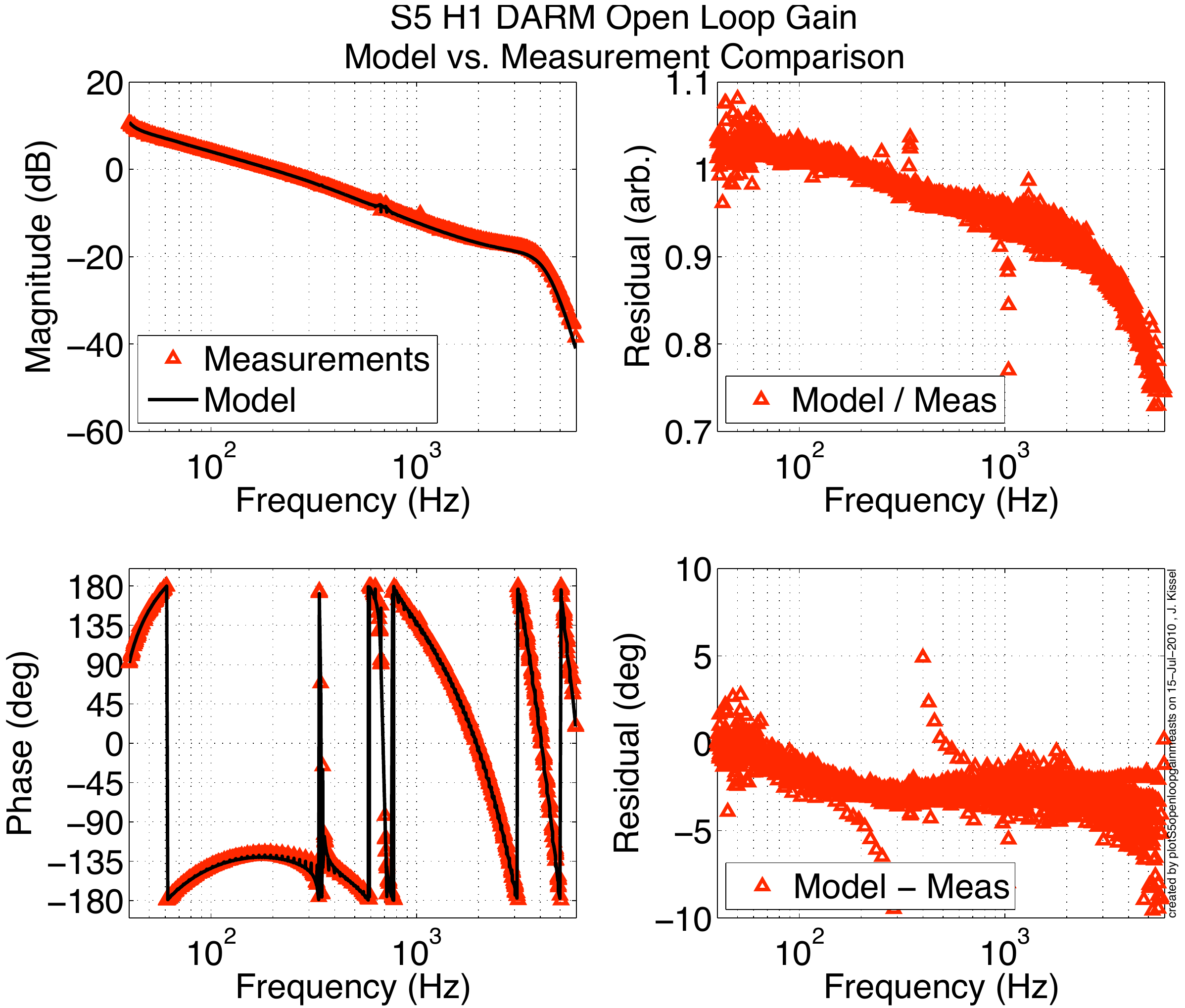}
  \end{minipage}
  \begin{minipage}{70mm}
\includegraphics[width=70mm]{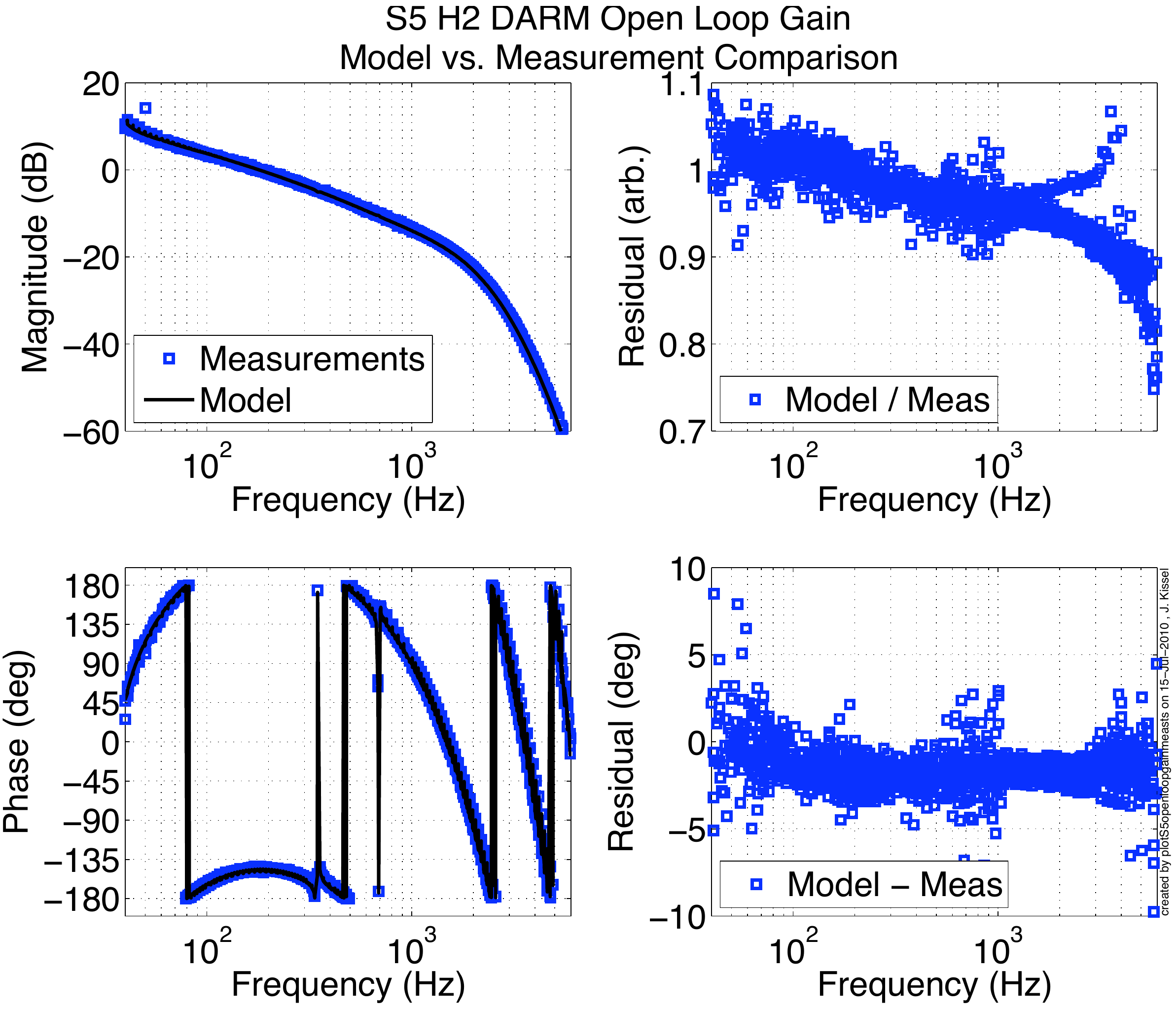}
  \end{minipage}
  \begin{center}
  \begin{minipage}{70mm}
\includegraphics[width=70mm]{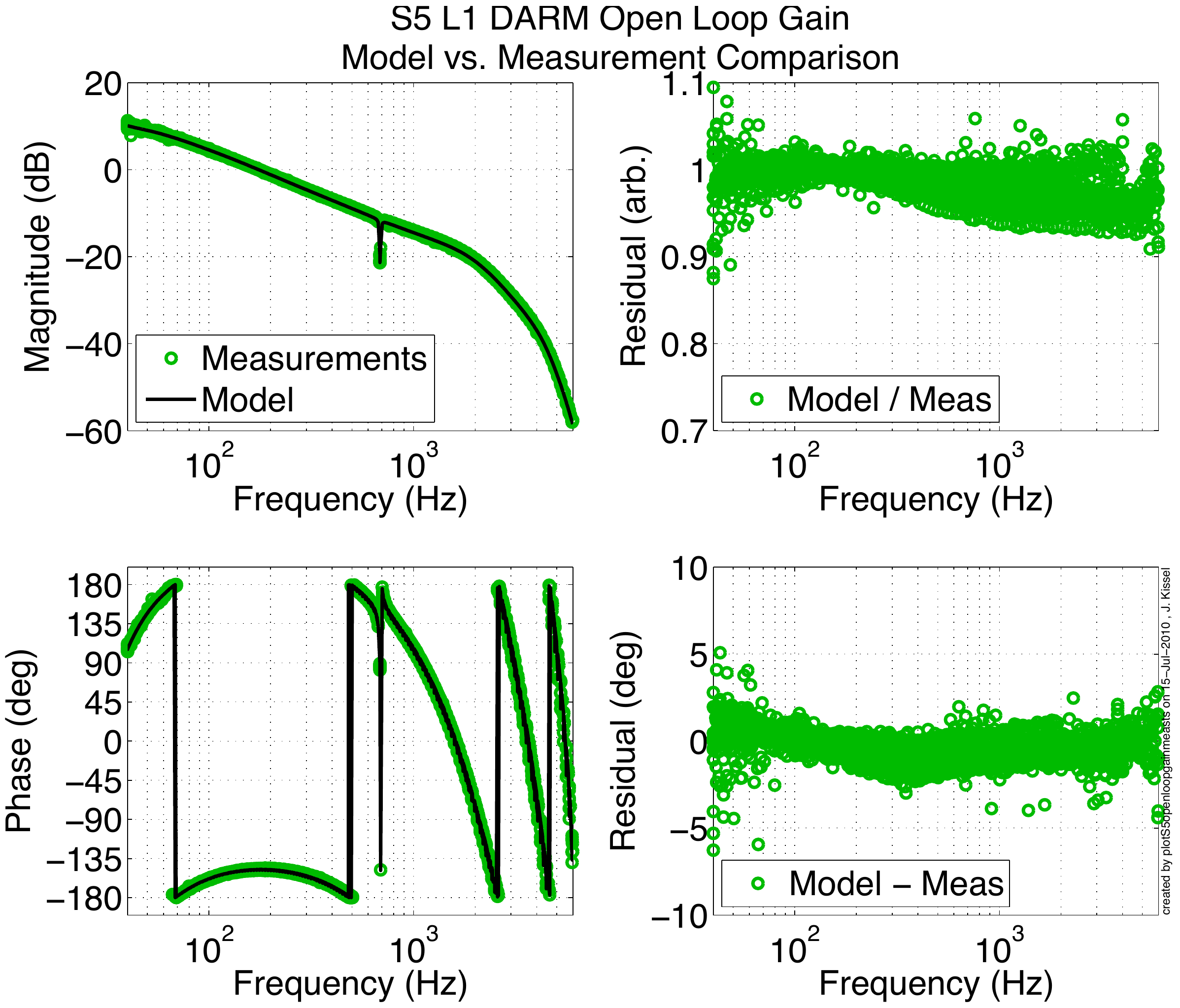}
  \end{minipage}
\end{center}
  \caption{Open loop transfer function model vs. measurement comparisons for H1 (top left), H2 (top right), and L1 (bottom) in all of S5. The four panels shown are the magnitude and phase of model and measurements (top and bottom left), and the ratio between model and measurements (top and bottom right). \label{olgplots}}
 \end{figure}

We measure the open loop transfer function many times during the course of the science run. To compare these measurements against the model for each epoch, they are normalized by the magnitude of the open loop transfer function at a fixed unity gain frequency. This normalization removes the time dependent scale factors between measurement times such that a fair comparison can be made. Figure \ref{olgplots} shows the results of this comparison.

The uncertainty estimation in the open loop transfer function magnitude and phase ($(\sigma_{|G_{L}|}/|G_{L}|)^{2}$ and $\sigma_{\phi_{G_{L}}}^{2}$) are separated into systematic and statistical uncertainty. We expect the ratio of the model and our measurements to follow a Gaussian distribution with unity mean in magnitude and zero mean in phase. This ratio is shown in Figure \ref{olgplots}. We observe a non-Gaussian systematic in all detectors from an unknown source, most apparent in the Hanford detectors. We estimate this systematic uncertainty in magnitude and phase by subtracting a smoothed version of the residuals, $G_{L}^{res}(f) = \langle G_{L}^{model} / G_{L}^{meas} \rangle$, from unity and zero, respectively. The statistical uncertainty, $\sigma_{\Sigma|G_{L}|}$ and $\sigma_{\Sigma \phi_{G_{L}}}$, is estimated from the standard deviation of the remaining scatter in the ratio after the systematic error $G_{L}^{res}(f)$ is subtracted. Both the systematic and statistical errors are added in quadrature to form the total uncertainty in the open loop transfer function model,
\begin{eqnarray}
\left(\frac{\sigma_{|G_{L}|}}{|G_{L}|}\right)^{2} & = & \left(\sigma_{\Sigma|G_{L}|}\right)^{2} + \left(1 - |G_{L}^{res}(f)|\right)^{2}  \\
\sigma_{\phi_{G_{L}}}^{2} & = &  \left(\sigma_{\Sigma \phi_{G_{L}}}\right)^{2} + \left(\phi_{G_{L}^{res}(f)}\right)^{2}.
\end{eqnarray}

\section{Uncertainty Estimation}
\label{error}
The measurement uncertainty of each component of the response function described in \S\ref{meas} are folded into a complex function of frequency known as the ``error budget.'' 

We do not assign any uncertainty to the digital filters $D(f)$ nor directly to the time-independent component of the sensing function $C_{L}(f)$. The digital filters, which are well-known digital functions, are placed into the model without uncertainty. As described in \S \ref{sensmeasts}, the frequency dependence of the sensing function is composed of parameters measured to negligible uncertainty. Uncertainties in its scaling coefficient $\mathcal{K}_{C}$, are accounted for in the open loop transfer function and actuation function uncertainty.

The uncertainties of the remaining quantities in the response function $A(f), G_{L}(f),$ and $\gamma(t)$ are treated as uncorrelated. If the uncertainties are completely correlated (i.e. there are none in $C_{L}(f)$), the covariant terms in the estimation reduce the overall estimate of the response function uncertainty \cite{resperror}. Since we do not have an independent estimate of the uncertainty in the sensing function, we adopt this conservative estimate.

We re-write the response function in terms of the measured quantities to which we assign uncertainty,
\begin{eqnarray}
R_{L}(f,t) = A(f)D(f) ~\frac{1 + \gamma(t)G_{L}(f)}{\gamma(t)G_{L}(f)}
\end{eqnarray}
and separate into magnitude and phase (dropping terms which include the uncertainty in $D(f)$),
\begin{eqnarray}
|R_{L}| & = &  \sqrt{\left(\frac{|A|}{\gamma |G_{L}|}\right)^{2} \Big[1 + (\gamma |G_{L}|)^{2} + 2 \gamma |G_{L}| \cos{(\phi_{G_{L}})}\Big]}, \label{magR}\\
\phi_{R_{L}} & = & \textrm{arctan}\left(\frac{\gamma |G_{L}| \sin{(\phi_{A})} + \sin{(\phi_{A} - \phi_{G_{L}})}}{\gamma |G_{L}| \cos{(\phi_{A})} + \cos{(\phi_{A} - \phi_{G_{L}} )}}\right), \label{phaR}
\end{eqnarray}
such that the relative uncertainty in magnitude and absolute uncertainty in phase are
\begin{eqnarray}
\left(\frac{\sigma_{|R_{L}|}}{|R_{L}|}\right)^{2}  \!\!\! & = & \!\!\! \left(\frac{\sigma_{|A|}}{|A|}\right)^{2}  +  \Re e \{W\}^{2} \left(\frac{\sigma_{|G_{L}|}}{|G_{L}|}\right)^{2} \nonumber\\
& & \hspace{1.2cm} ~+~ \Im m\{W\}^{2} \sigma_{\phi_{G_{L}}}^{2} + \Re e \{W\}^{2} \left(\frac{\sigma_{\gamma}}{\gamma}\right)^{2} \label{magerror} \\
& & \nonumber\\
\sigma_{\phi_{R_{L}}}^{2} & = & \sigma_{\phi_{A}}^{2} + \Im m \{W\}^{2} \left(\frac{\sigma_{|G_{L}|}}{|G_{L}|}\right)^{2} \nonumber\\
& & \hspace{0.5cm} ~+~ \Re e\{W\}^{2} ~\sigma_{\phi_{G_{L}}}^{2} + \Im m \{W\}^{2} \left(\frac{\sigma_{\gamma}}{\gamma}\right)^{2}, \label{phaerror}
 \end{eqnarray}
where we define $W \equiv 1 / (1 + G_{L})$ \cite{resperror}. Each uncertainty component in Eqs. \ref{magerror} and \ref{phaerror} is assumed to be the same over the course of the science run (independent of epochs).  However, the complex coefficient $W$ is different for each epoch. 

\begin{figure}[h!]
\begin{center}
\includegraphics[width=90mm]{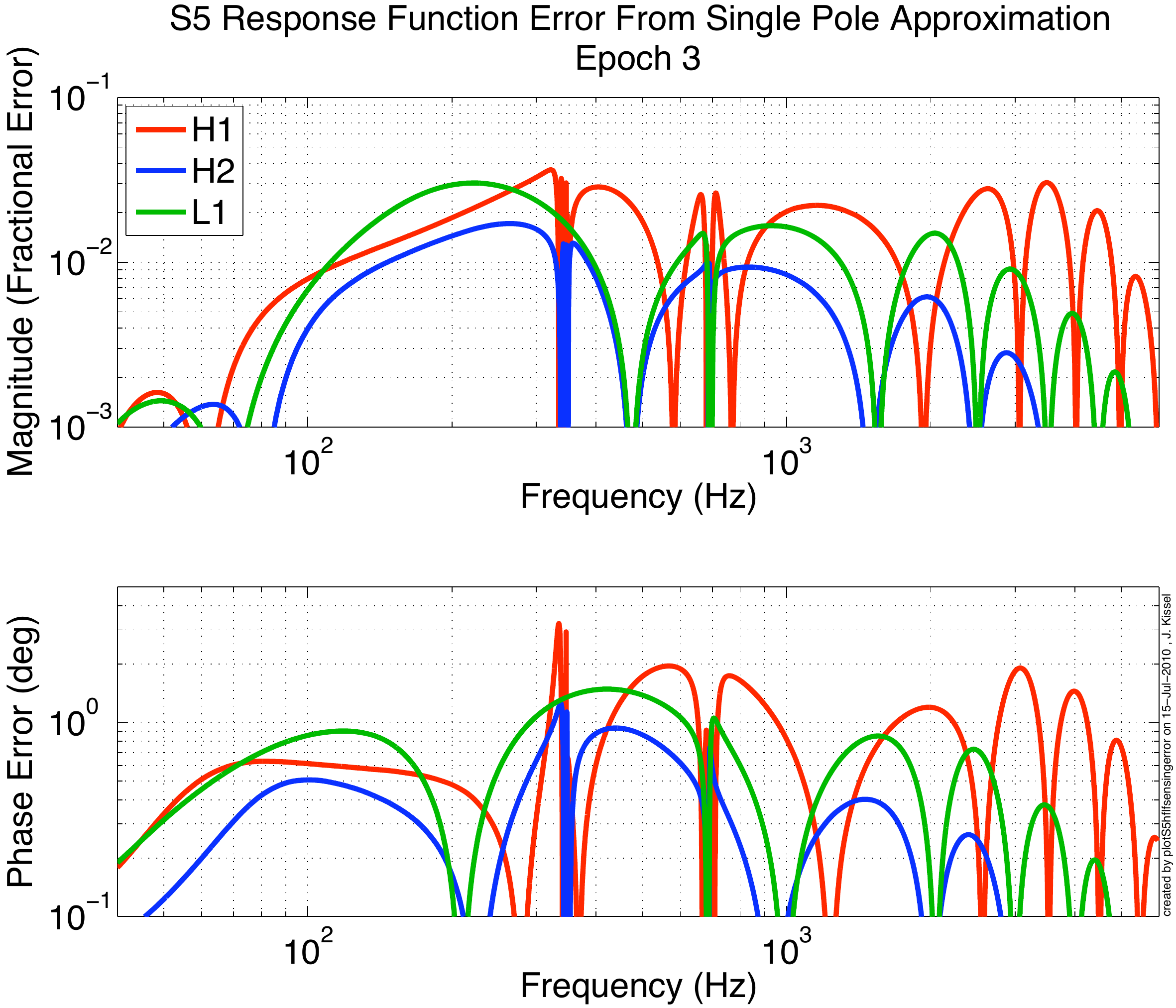}
\end{center}
\caption{Systematic uncertainty in response function arising from the single pole approximation of the Fabry-Perot cavity response in open loop transfer function. This uncertainty is epoch-dependent; only the third epoch for each detector is shown.  \label{hflferror}}
\end{figure}

Our calculation of the response function includes the open loop transfer function model which is approximated by replacing the complete cavity response $H_{FP}(f)$ (Eq. \ref{FP}) with the single pole transfer function $H_{SP}(f)$ (Eq. \ref{SP}) in the sensing function subsystem. We include the ratio of the response function calculated with and without the correct cavity response in our error budget,
\begin{eqnarray}
\frac{R_{L}^{FP}(f)}{R_{L}^{SP}(f)} & = & \frac{1 + (H_{FP}/H_{SP})  G_{L}(f)}{1 + G_{L}(f)}
\end{eqnarray}
added linearly (as opposed to in quadrature) because the approximation results in a frequency-dependent scaling of the response function with known sign. As with the weighting function $W$, this term involves the direct multiplication of the open loop transfer function and therefore is epoch dependent. As an example, Figure \ref{hflferror} shows this error contribution from the third epoch in each detector.

\section{Results}
\begin{figure}[h!]
  \begin{minipage}{70mm}
\includegraphics[width=70mm]{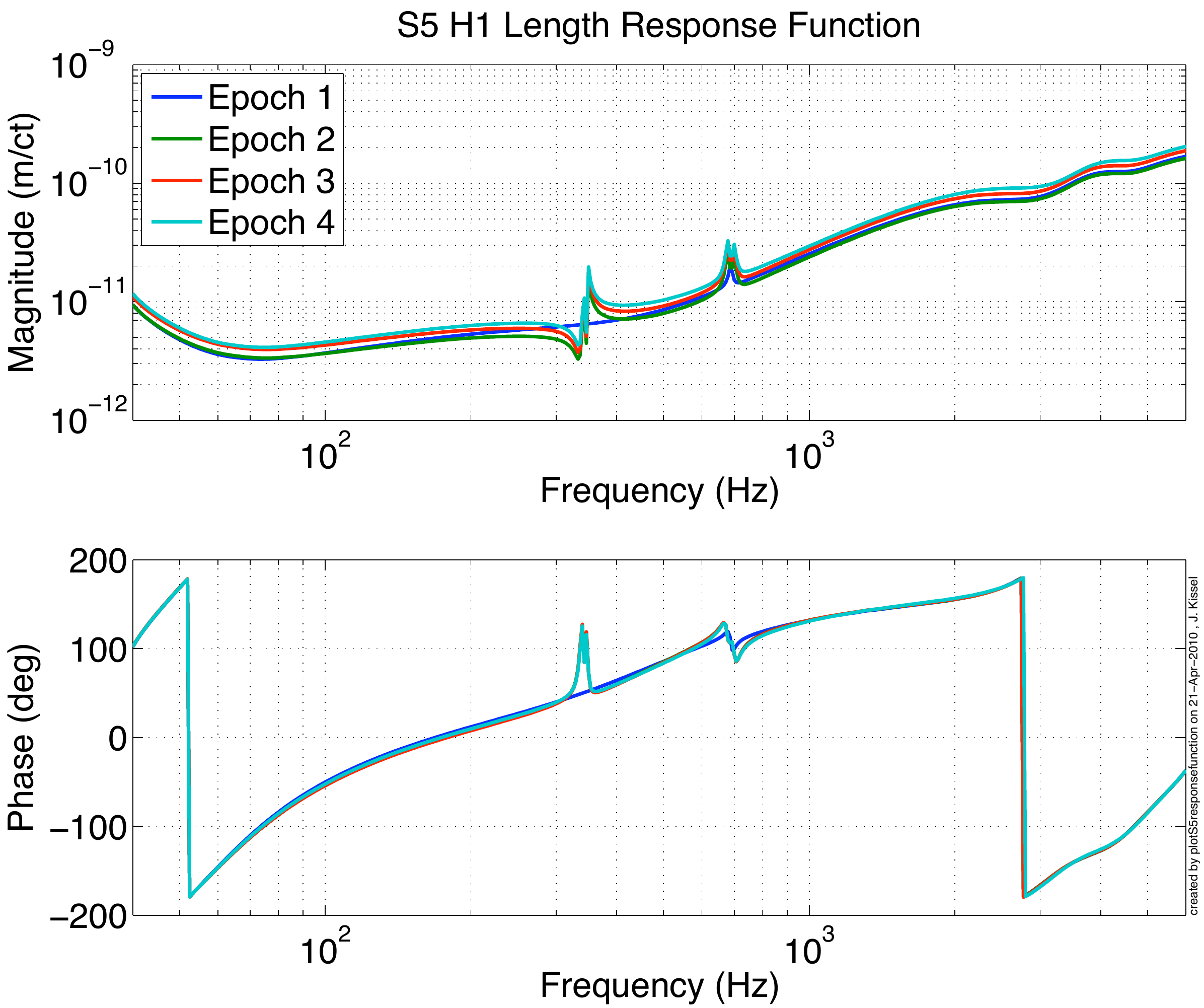}
  \end{minipage}
  \begin{minipage}{70mm}
\includegraphics[width=70mm]{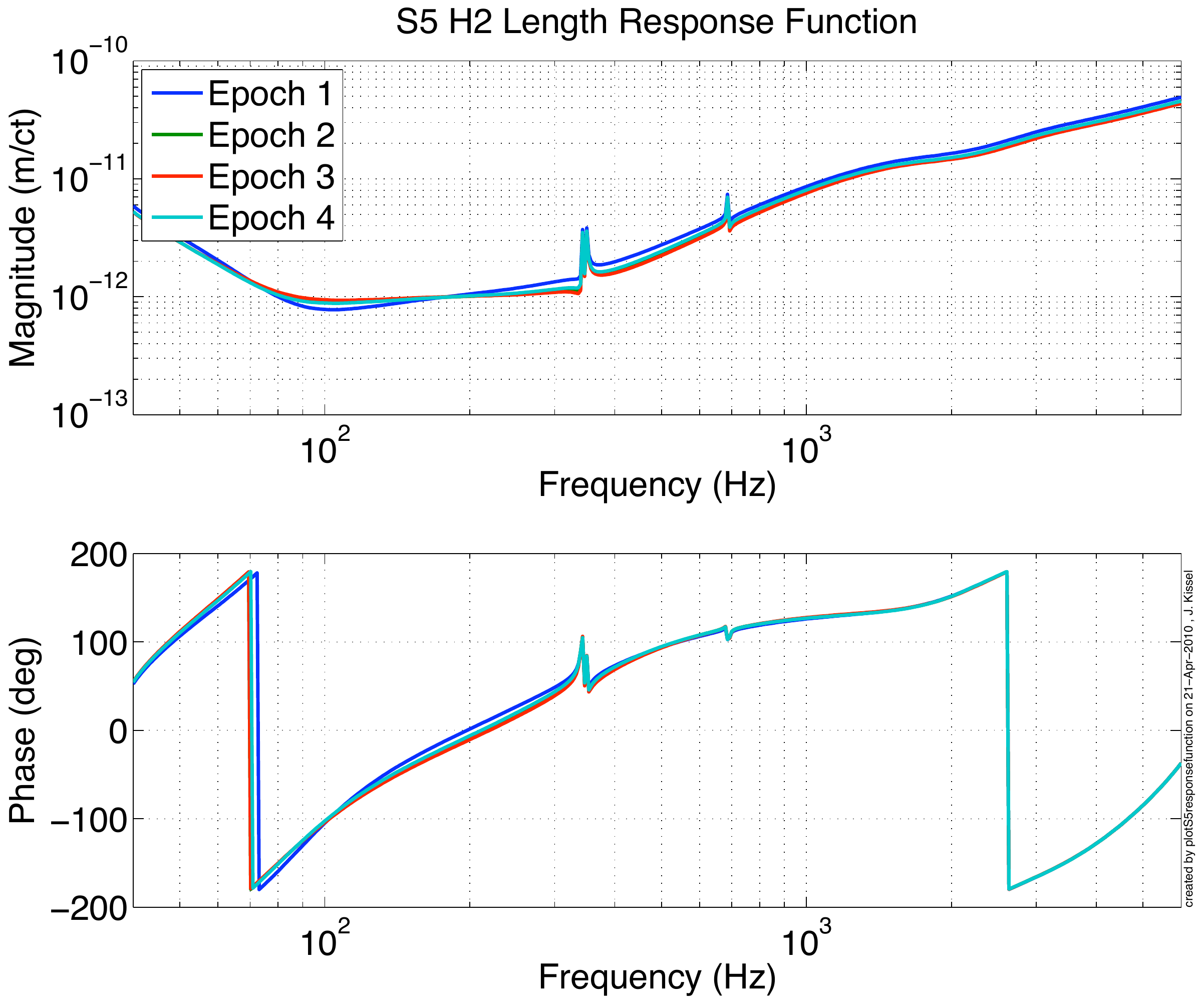}
  \end{minipage}
 \begin{center}
  \begin{minipage}{70mm}
\includegraphics[width=70mm]{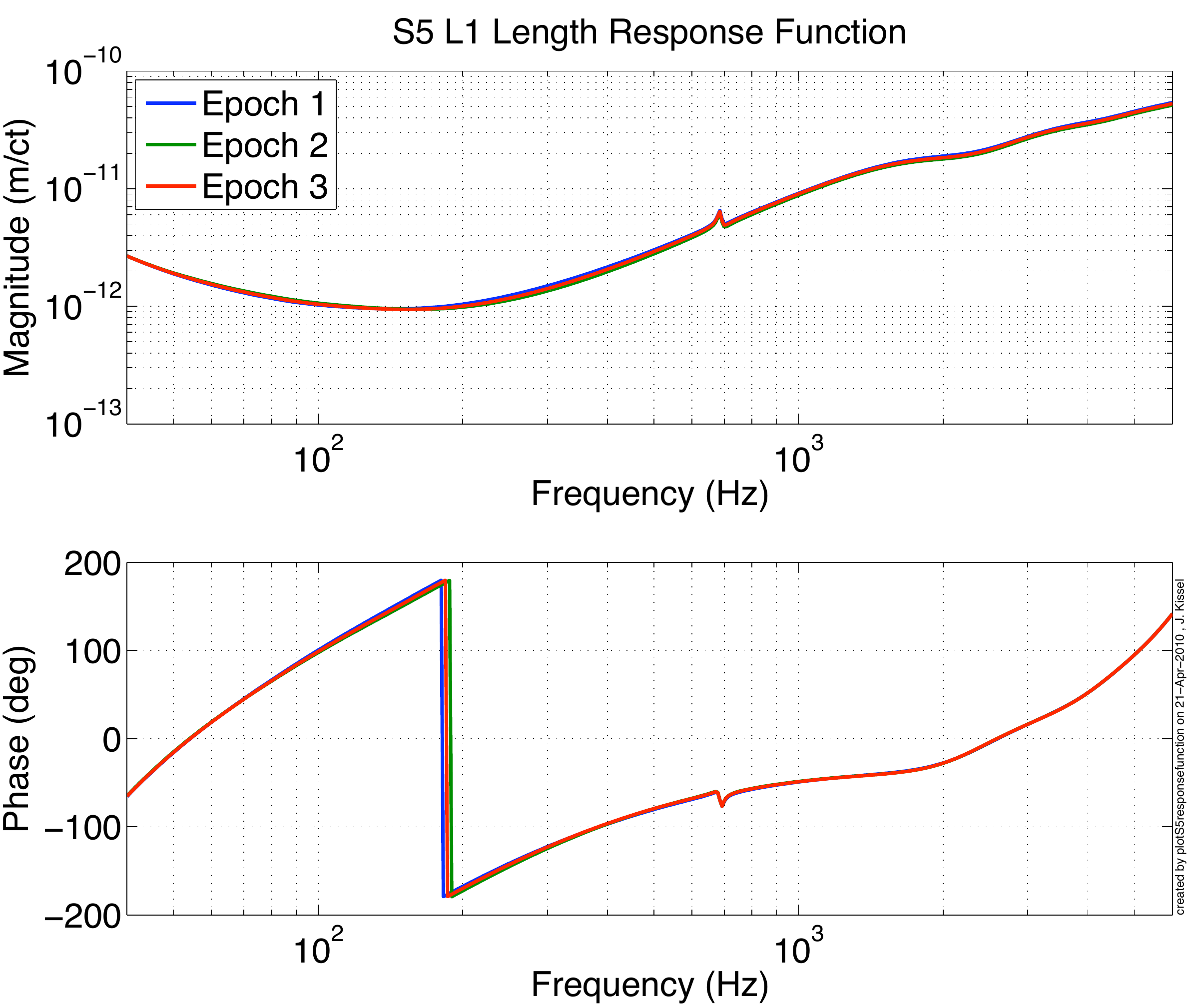}
  \end{minipage}
  \end{center}
  \caption{Frequency dependent response function, $R_{L}(f)$, for the three LIGO interferometers for all epochs of the S5 science run. \label{finalrespfunction}}
  \end{figure}

\begin{figure}[h!]
  \begin{minipage}{70mm}
\includegraphics[width=65mm]{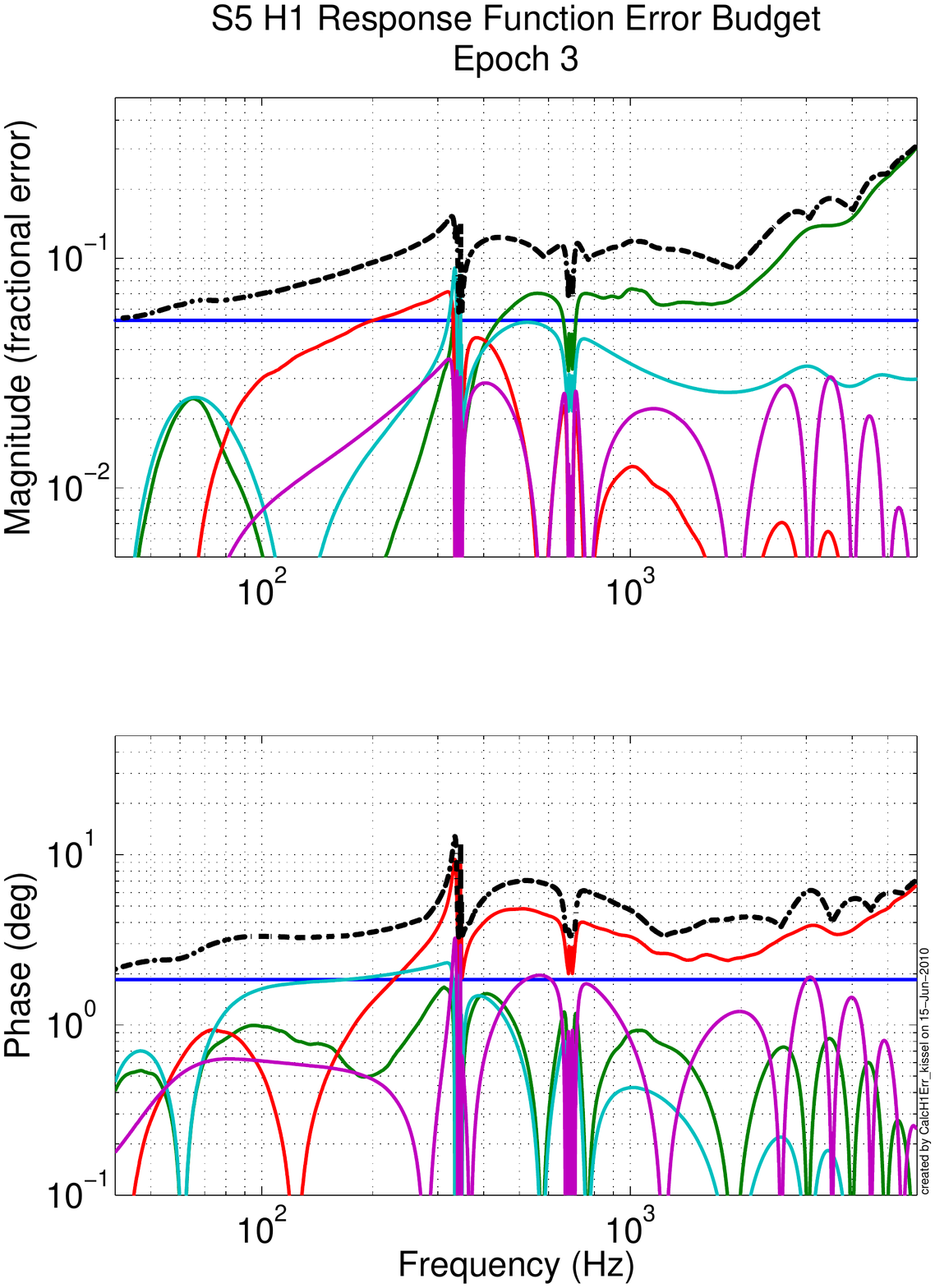}
  \end{minipage}
  \begin{minipage}{70mm}
\includegraphics[width=65mm]{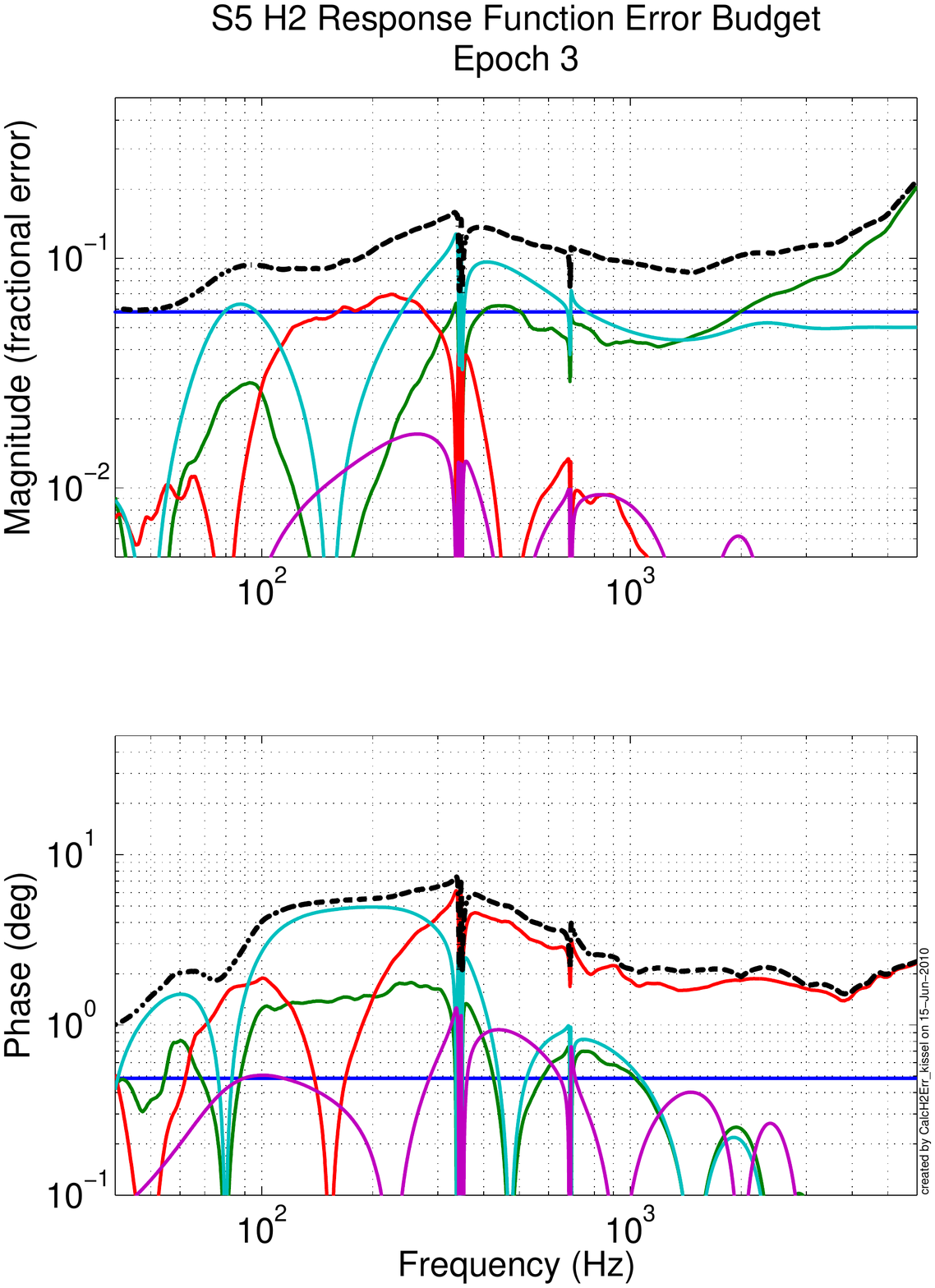}
  \end{minipage}
  \begin{minipage}{70mm}
 \includegraphics[width=65mm]{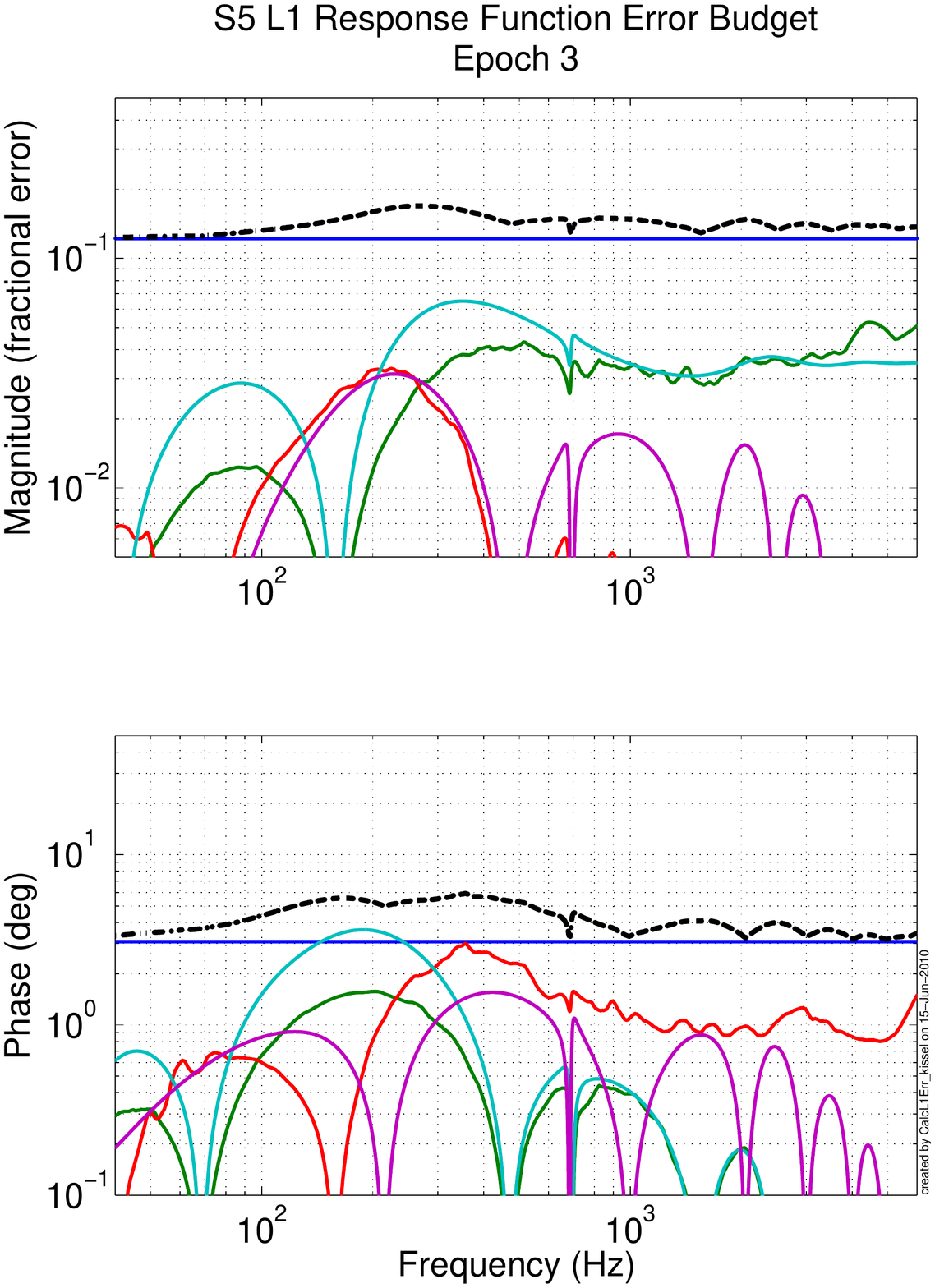}
  \end{minipage}
  \begin{minipage}{35mm}
  \begin{center}
\includegraphics[width=35mm]{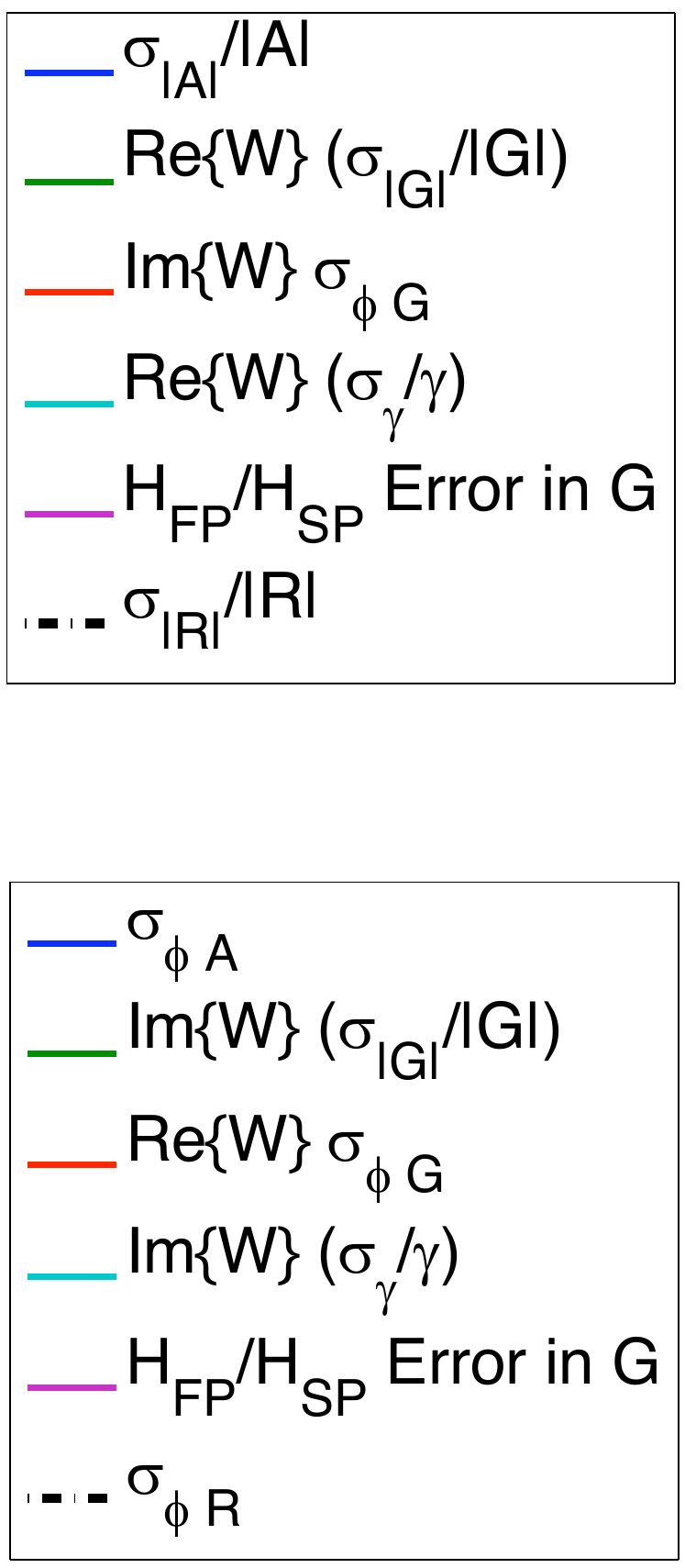}
\end{center}
  \end{minipage}
  \caption{Frequency dependent error estimates for the response function for H1 (top left), H2 (top right), and L1 (bottom left). In magnitude (top panel) and phase (bottom panel), the total uncertainty (dashed-black) is composed of the uncertainty in actuation, $\sigma_{|A|}/|A|, ~\sigma_{\phi_{A}}$ (in blue), the open loop transfer function magnitude $\sigma_{|G|}/|G|$ (in green), the open loop transfer function phase, $\sigma_{\phi_{G}}$ (in red), the time-dependent factor, $\sigma_{\gamma}/\gamma$ (in cyan), and from the single pole approximation of the Fabry-Perot cavity response (in magenta).  \label{toterror}}
  \end{figure}

In Figure \ref{finalrespfunction} we plot the final response function for all interferometers for the entire fifth science run. Figure \ref{toterror} shows the frequency dependence of all terms in the error budget of the response function for the third epoch of each detector. In Table \ref{tableoferrors}, we summarize the frequency-dependent uncertainty of each interferometer's response function by dividing the error into three frequency bands: 40-2000 Hz, 2000-4000 Hz and 4000-6000 Hz and computing the RMS errors across each band, averaged over all epochs. All epoch uncertainties are within 1\% of the mean uncertainty stated.

 \begin{table}[h!]
\caption{Summary of band-limited response function errors for the S5 science run. \label{tableoferrors}}
\begin{center}
\begin{tabular}{|c|c|c|c||c|c|c|}
\hline
\multirow{3}{*}{{\bf }} & \multicolumn{3}{|c||}{{\bf $R_{L}(f)$ Magnitude Error (\%)}} & \multicolumn{3}{c|}{{\bf $R_{L}(f)$ Phase Error (Deg)}}\\ 
& &  &  &  &  & \\
 & 40-2000 Hz & 2-4 kHz & 4-6 kHz & 40-2000 Hz & 2-4 kHz & 4-6 kHz \\ \hline
H1 & 10.4 & 15.4 & 24.2  & 4.5 & 4.9 & 5.8 \\
H2 & 10.1 & 11.2 & 16.3  & 3.0 & 1.8 & 2.0 \\
L1 & 14.4 & 13.9 & 13.8 & 4.2 & 3.6 & 3.3 \\
\hline
\end{tabular}
\end{center}
\end{table}

The largest source of systematic error in most data analysis techniques used to analyze S5 LIGO data is the uncertainty in response function magnitude \cite{pulsar, burst,stoch}. Our inability to measure the sensing function independently of the closed loop (specifically its scaling coefficient) forces a conservative, uncorrelated treatment of the uncertainty in the measured subsystems, $A(f)$ and $G_{L}(f)$, inflating the total uncertainty in the response function. In all detectors, we find the uncertainty in the actuation function, $A(f)$, dominates the response function error budget in magnitude. 

The statistical uncertainty in the free-swinging Michelson measurements of the actuation scaling coefficient are the primary source of the actuation uncertainty. In the Hanford detectors, the uncertainty arises from our inability to displace the test mass above residual external noise sources at high frequency. This decreases the signal-to-noise of the measurement, inflating the uncertainty estimate across the measurement band. For L1, in which we have obtained a large number of measurements using several methods of the free-swinging Michelson technique (see \ref{simpmichappendix}), we have found the results to be inconsistent with a Gaussian distribution. We attribute this to a poorly understood underlying variation in the technique, for example the assumption that the optical gain is time-independent over the course of the measurement suite. 

The assumption that the actuation scaling coefficient is linear in amplitude over the range of actuation, from the $10^{-8}$ m employed for the free-swinging Michelson technique to the $10^{-18}$ m required to compensate for expected gravitational wave signals, has not been confirmed. To investigate the linearity of the actuation scaling coefficients over this range of actuation amplitudes, and to bound potential overall systematic errors, we have employed two additional, fundamentally different, actuator calibration methods. The so-called ``frequency modulation'' technique \cite{VCO} uses an independently calibrated oscillator to frequency-modulate the interferometer's laser light, creating an effective length modulation on the order of $10^{-13}$ m while operating in a single-arm interferometer configuration. The so-called ``photon calibrator'' technique \cite{pcal} uses auxiliary, power-modulated lasers to displace the test masses by approximately $10^{-18}$ m via radiation pressure with the interferometer in its nominal configuration (see Figure 1). Both methods are employed at select frequencies across the LIGO measurement band. Statistical uncertainties for both methods are reduced to the 1\% level by averaging many measurements. 

At the end of the S5 science run, a detailed comparison between these two methods and the free-swinging Michelson technique was performed. With all three calibration methods, actuation coefficients were measured over the frequency band from 90~Hz to 1~kHz for each end test mass. For the H1 and H2 interferometers, all calculated actuation coefficients--for all frequencies, for all four masses, and for all three methods--were within a $\pm15$\% range. The maximum difference between the mean value for any method and the mean value for all three methods, for any of the four end test masses, was 3.7\% \cite{dccalcomp}. This indicates that the overall systematic uncertainties in the actuation functions determined using the free-swinging Michelson method, and therefore the magnitudes of the interferometer response functions, are within these bounds.

\clearpage
\newpage

\section{Summary}
\label{summary}
The LIGO interferometers have provided the some of the world's most sensitive gravitational wave strain measurements during their fifth science run. We have described a model used for each interferometer's differential arm length control loop known as the length response function, $R_{L}(f,t)$, the proportionality between the digital Pound-Drever-Hall error signal and differential displacement of the end test masses. Measurements presented here have shown the frequency-dependent uncertainty in $R_{L}(f,t)$ is less than 15\% in magnitude and $5^{\circ}$ in phase in the frequency band where the interferometer is most sensitive. Because we cannot measure the sensing function without the interferometers under control, this estimate is limited by our ability to measure the actuation function. The results of two fundamentally different, high-precision methods for measuring the actuation functions \cite{VCO,pcal} confirm that the free-swinging Michelson results are within the stated uncertainties \cite{dccalcomp}.

In the two calendar year science run, as our knowledge of the long-term characteristics of the instrument increased, a great deal of improvements were made to our measurement techniques compared with prior results \cite{S1cal}.  However, future detectors will have more sophisticated actuation and sensing methods \cite{2ndGen,aLIGO}. In addition, an amplitude uncertainty of 10\% or less is required to reduce the calibration uncertainty below other systematic errors in the continually improving astrophysical searches \cite{CalAcc}. To achieve this goal, the non-Gaussian distribution of the actuation function measurements must be better understood and independent techniques of measuring the actuation coefficient, like the frequency modulation and photon calibrator, must be used in concert with the standard techniques presented in this paper to reduce limiting systematic errors. 

\section*{Acknowledgments}
The authors gratefully acknowledge the support of the United States
National Science Foundation for the construction and operation of the
LIGO Laboratory and the Science and Technology Facilities Council of the
United Kingdom, the Max-Planck-Society, and the State of
Niedersachsen/Germany for support of the construction and operation of
the GEO600 detector. The authors also gratefully acknowledge the support
of the research by these agencies and by the Australian Research Council,
the Council of Scientific and Industrial Research of India, the Istituto
Nazionale di Fisica Nucleare of Italy, the Spanish Ministerio de
Educacion y Ciencia, the Conselleria d'Economia Hisenda i Innovacio of
the Govern de les Illes Balears, the Royal Society, the Scottish Funding 
Council, the Scottish Universities Physics Alliance, The National Aeronautics
and Space Administration, the Carnegie Trust, the Leverhulme Trust, the David
and Lucile Packard Foundation, the Research Corporation, and the Alfred
P. Sloan Foundation.

LIGO was constructed by the California Institute of Technology and 
Massachusetts Institute of Technology with funding from the National Science Foundation 
and operates under cooperative agreement PHY-0107417. This paper has LIGO Document Number LIGO-P0900120.

 \appendix

 \section{The Free-Swinging Michelson Techniques}
 \label{simpmichappendix}
The technique used for determining the actuation coefficients, $\mathcal{K}_{A}^{x,y}$, for the fifth science run is known as the ``free-swinging Michelson'' technique. This technique uses the interferometer's well-known Nd:YaG laser wavelength ($\lambda = 1064.1 \pm 0.1$ nm, \cite{lambda1,lambda2}) as the calibrated length reference while using the test mass coil actuators to cause a change in length of simple interferometer configurations. The technique may be used in two similar methods: the ``Simple Michelson'' and ``Asymmetric Michelson'' methods. 

\begin{figure}[h!]
\begin{center}
\includegraphics[width=140mm]{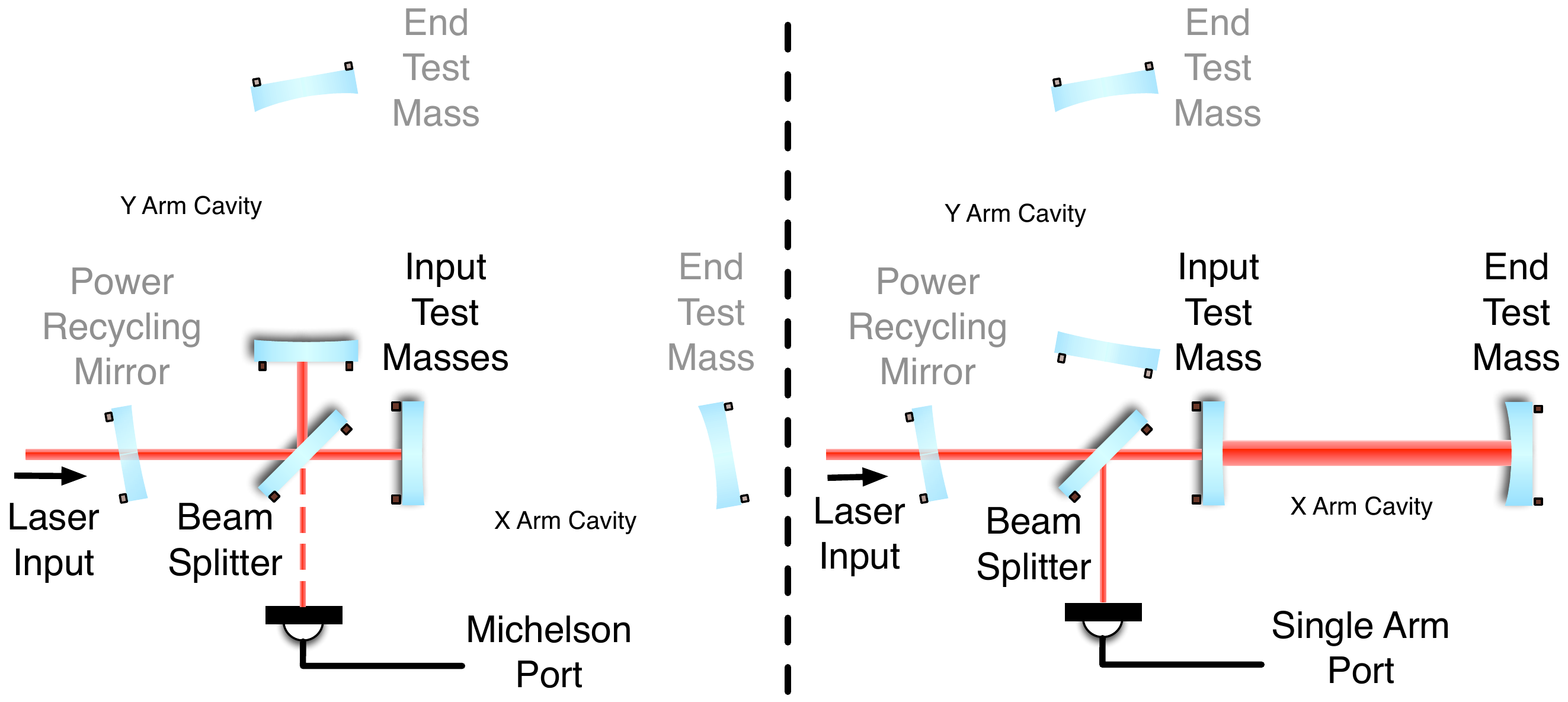}
\end{center}
\caption{Interferometer configurations used during the simple Michelson method of measuring the actuation scaling coefficient. Left: The simple Michelson configuration, where the power recycling mirror and end test masses are misaligned. Right: The single arm configuration, with the power recycling mirror and the opposing arm's input and end test mass are misaligned. \label{simplemich}}
\end{figure}

The Simple Michelson method is composed of two steps. The first step determines the actuation scaling coefficient for the input test masses $\mathcal{K}_{i}$ with the interferometer in a non-standard configuration called a frequency-modulated simple Michelson (see left panel of Figure \ref{simplemich}). The second step determines the end test mass actuation coefficient, $\mathcal{K}_{A}$, from the input test coefficient,  $\mathcal{K}_{i}$, and transfer function measurements of the input and end test masses of a single Fabry Perot arm cavity (see right panel of Figure \ref{simplemich}). The Asymmetric Michelson determines $\mathcal{K}_{A}$ directly using the configuration shown in Figure \ref{asymmich}. Both free-swinging Michelson methods are described below.
 
\subsection*{Simple Michelson Method}
The Pound-Drever Hall error signal at the anti-symmetric port, $q_{AS}$, for a frequency-modulated simple Michelson interferometer is
\begin{eqnarray}
q_{AS} & = & \frac{1}{2} A_{pp} ~\sin\left(\frac{4 \pi}{\lambda} ~\Delta \ell \right),
\end{eqnarray}
where $A_{pp}$ is the peak-to-peak amplitude of the signal (proportional to the input power, the product of Bessel functions of modulation strength, and the transmission of the sidebands into the antisymmetric port from the Michelson asymmetry), $\lambda = 1064$ nm is the wavelength of the input laser light,  and $\Delta \ell = \ell_{x} - \ell_{y}$ is the differential arm length of the Michelson. $A_{pp}$ is measured by aligning the simple Michelson and recording the $q_{AS}$ time series as it is left uncontrolled. In this configuration, external noise sources (e.g. residual ground motion) are large enough to cause the Michelson to sweep through many interference fringes.

For the simple Michelson, when $\Delta \ell/\lambda \ll 1$,  
\begin{eqnarray}
q_{AS} & \approx & k \Delta \ell 
\end{eqnarray} 
with the simple Michelson's ``optical gain,'' 
\begin{eqnarray}
k ~=~ (2\pi / \lambda) A_{pp},  \label{opgain}
\end{eqnarray}
which has units of digital signal counts per meters of input test mass motion. After a measurement of $A_{pp}$ is obtained, we control the optics using their coil actuators, forcing the Michelson into the linear regime where Eq. \ref{opgain} is valid. 

\begin{figure}[h!]
\begin{center}
\includegraphics[width=90mm]{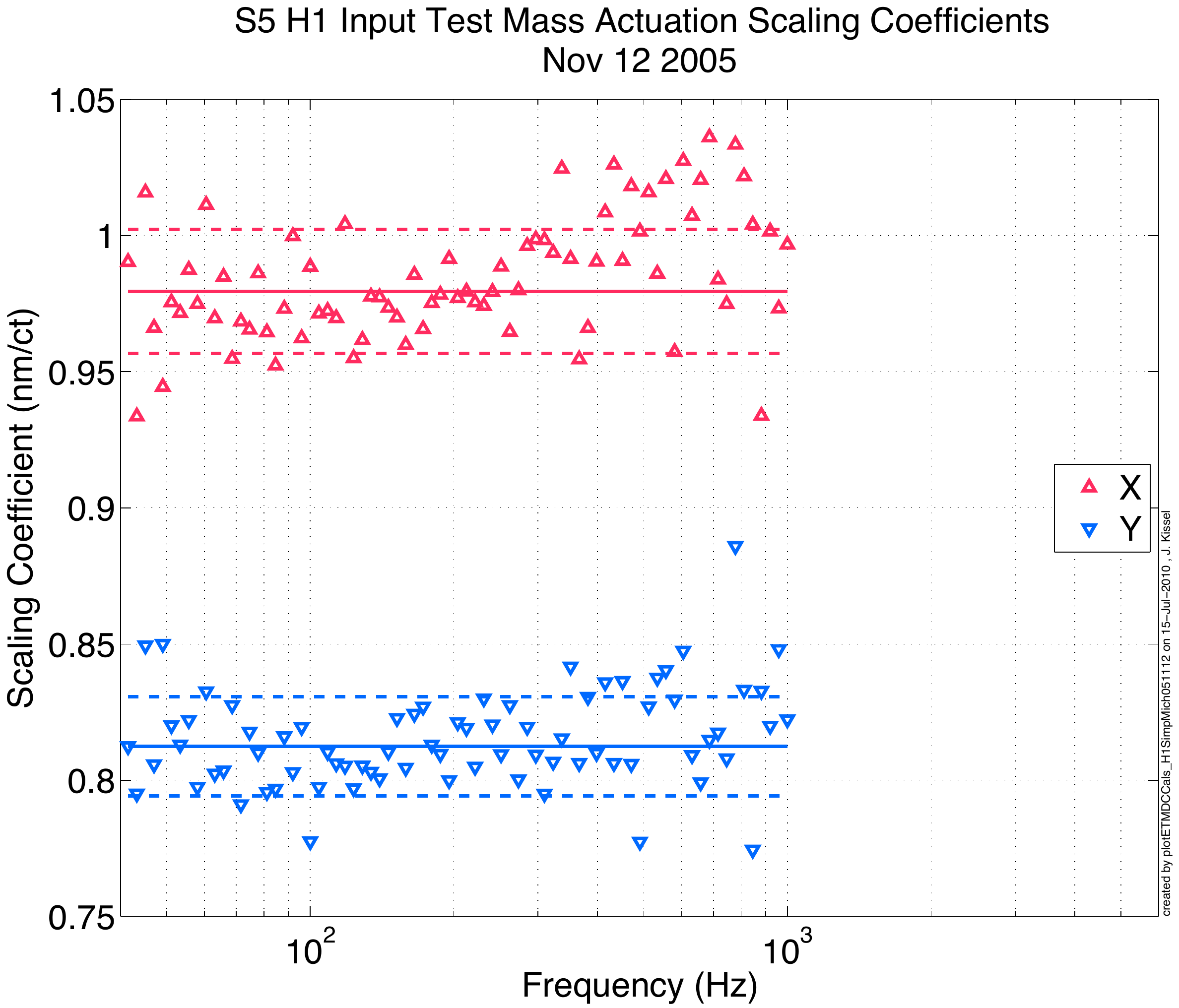}
\end{center}
\caption{Example actuation scaling coefficients for the H2 input test masses $\mathcal{K}_{i}^{x,y}$, measured using the simple Michelson method. Top: $\mathcal{K}_{i}$ as a function of frequency for the X (red) and Y (blue) input test mass. Solid lines indicate the median of the data points $\langle \mathcal{K}_{i} \rangle$, dashed lines indicate $1\sigma$ error bars (the standard deviation of all frequency points). \label{itmcal}}
\end{figure}

The actuation function of the suspended input test masses can be approximated by the center-of-mass force-to-displacement transfer function, $P_{cm}^{i}$ with a scaling coefficient, $\mathcal{K}_{i}$. We obtain a measurement of $\mathcal{K}_{i}$ for a given input test mass by introducing a digital excitation $exc_{i}$ into the control loop that is much larger than residual external noise sources. The excitation is performed over many frequencies in the gravitational wave band; assuming the model is complete, the coefficient should be frequency-independent across the band. We obtain a solution for the digital excitation counts on the input test mass in terms of meters of resulting motion as measured by  $q_{AS}$ (normalized by the pendulum response $P_{cm}^{i}$),
\begin{eqnarray}
\mathcal{K}_{i} & = & \left(\frac{q_{AS}}{exc_{i}}\right) \left(\frac{1 + G_{SM}}{k}\right) \left(\frac{1}{P_{cm}^{i}}\right) .
\end{eqnarray}
The first term is the measured response of the Michelson during the single input test mass excitation. The second term contains  the open loop transfer function $G_{SM}$ of the simple Michelson control loop (measured just prior to measuring the response to excitation) and the quantity $k$ is as defined in Eq. \ref{opgain}. We take the median of $\mathcal{K}_{i}$ (denoted with ``bra''``kets,'' $\langle~\rangle$), over the measured frequency points to remove measurement outliers and residual frequency dependence (or time dependence of $k$, as discussed in \S \ref{error}). Figure \ref{itmcal} shows an example measurement of $\mathcal{K}_{i}$ for each input test mass in H2.

\begin{figure}[h!]
\begin{center}
\includegraphics[width=90mm]{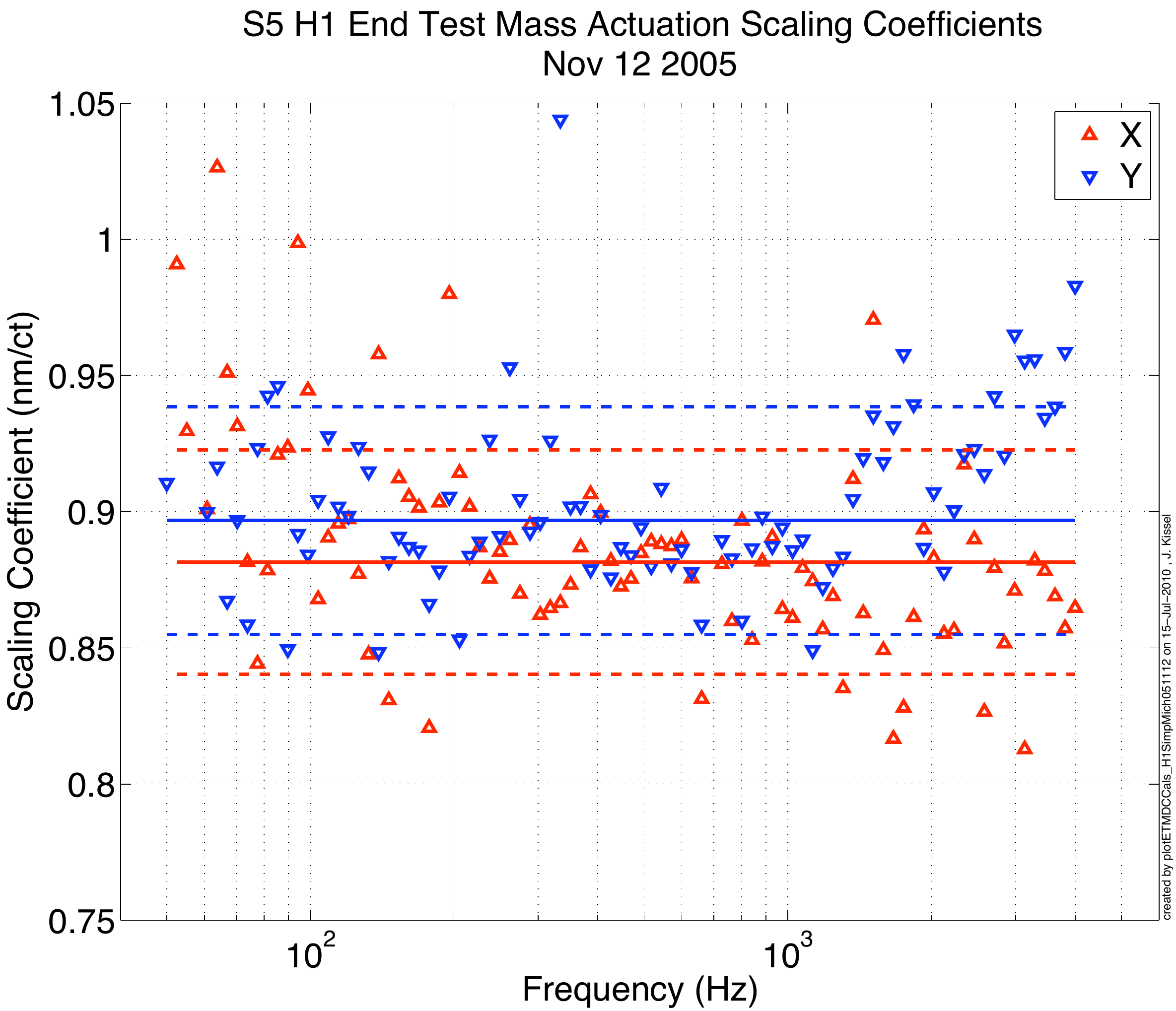}
\end{center}
\caption{Example actuation scaling coefficients for the H1 end test masses  $\mathcal{K}_{A}^{x,y}(f)$, measured using the simple Michelson method. Top: $\mathcal{K}_{A}$ as a function of frequency for the X (red) and Y (blue) end test mass. Solid lines indicate the median of the data points $\langle \mathcal{K}_{A}^{x,y}\rangle$, dashed lines indicate $1\sigma$ error bars (see \S \ref{error} for description). \label{oldmethod}}
\end{figure}

We then configure the interferometer to form a single Fabry-Perot cavity composed of one arm of the interferometer, and control it such that the cavity is under resonance (see Figure \ref{simplemich}). In this configuration, the response of the single arm cavity (now recorded by the in-phase demodulated output $i_{AS}$, see \cite{freqresp} for details) to sequential length excitations of the input test mass, $exc_{i}$, and end test mass, $exc_{e}$, are measured. The ratio, $\mathcal{R}_{ie}$ of these to transfer functions can then be used to write the actuation coefficient for the end test masses as
\begin{eqnarray}
\mathcal{K}_{A} & = & \mathcal{R}_{ie} \left(\frac{P_{cm}^{i}}{P_{cm}}\right) \langle \mathcal{K}_{i} \rangle ~=~ \left(\frac{i_{AS}}{exc_{e}} \right) \left(\frac{exc_{i}}{i_{AS}}\right) \left(\frac{P_{cm}^{i}}{P_{cm}}\right) \langle \mathcal{K}_{i} \rangle, \label{simpmich2}
\end{eqnarray}
where $\mathcal{K}_{A}$ has units of test mass motion in meters (as measured by $q_{AS}$) per count of digital excitation. Figure \ref{oldmethod} shows a measurement of $\mathcal{K}_{A}$ for each test mass in H1. As in the first step, the median of the frequency points measured in $\mathcal{K}_{A}$ is used to form a single value for the coefficient over the measurement bands. 

\subsection*{Asymmetric Michelson Method}
\begin{figure}[h!]
\begin{center}
\includegraphics[width=70mm]{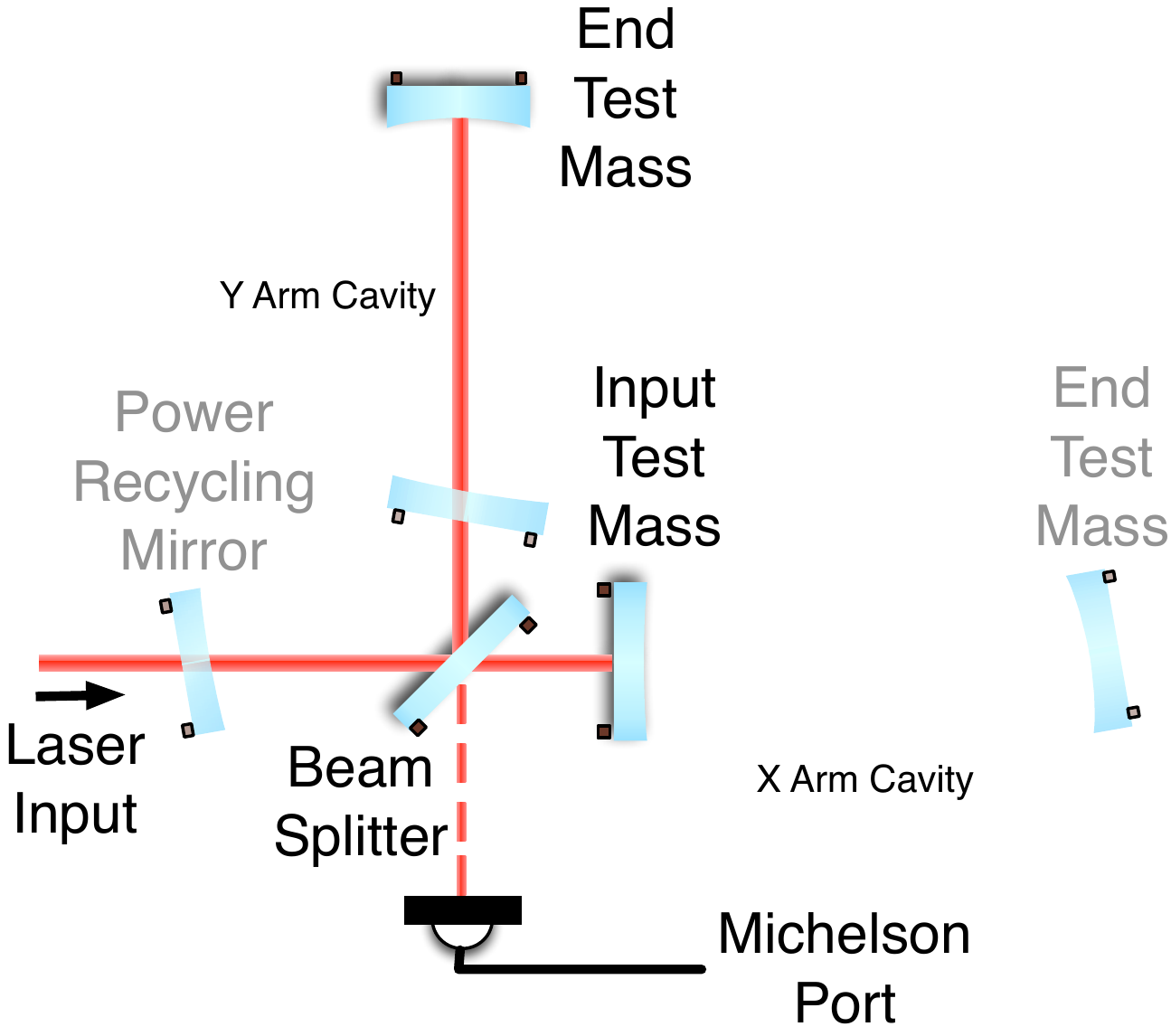}
\end{center}
\caption{``Asymmetric Michelson'' configuration of the interferometer. With the power recycling mirror misaligned, an input test mass and opposing end test mass are aligned. \label{asymmich}}
\end{figure}

\begin{figure}[h!]
\begin{center}
\includegraphics[width=90mm]{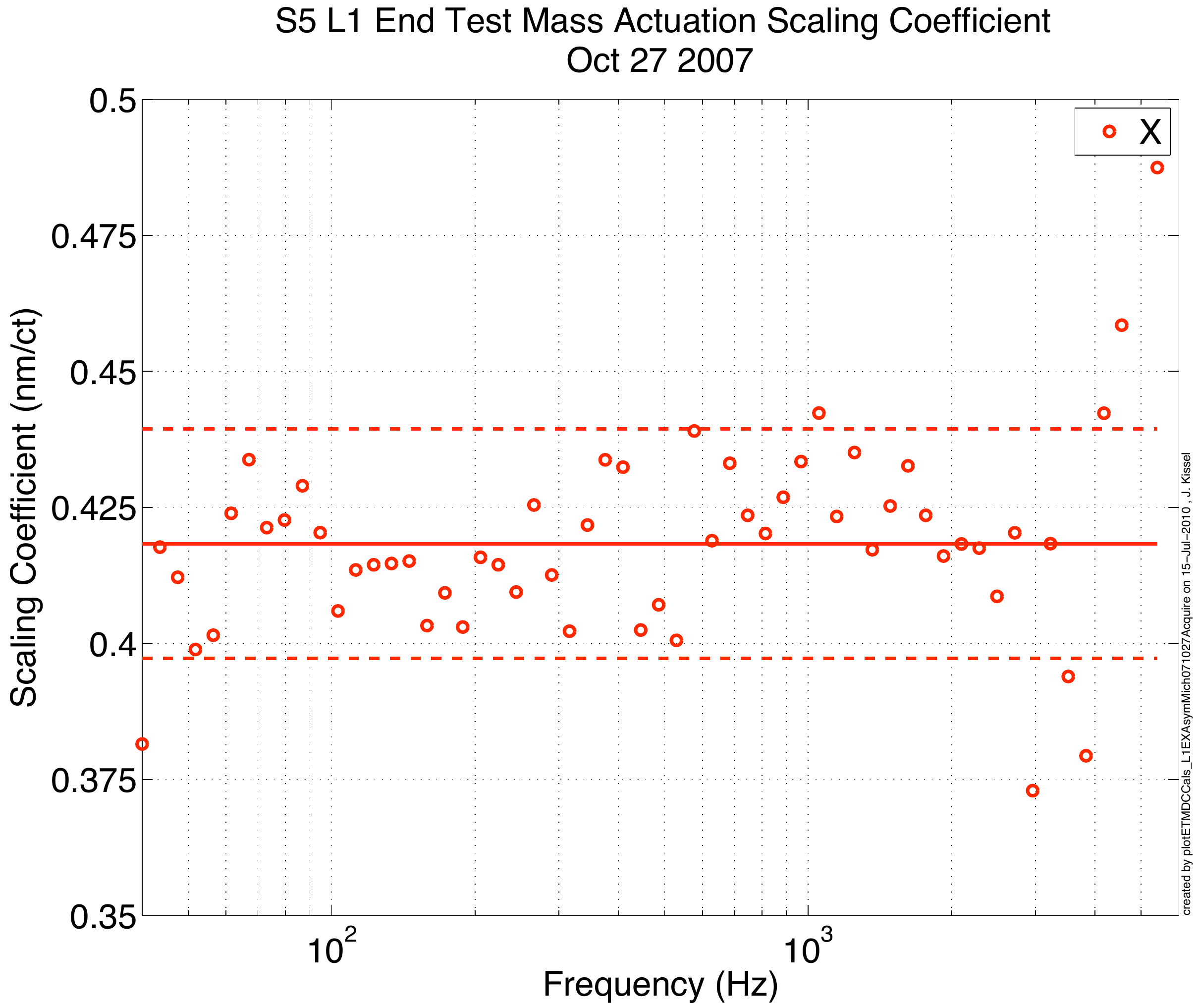}
\end{center}
\caption{Example actuation scaling coefficient for the L1 X arm end test mass  $\mathcal{K}_{A}^{x}$, measured using the asymmetric Michelson method. Top: $\mathcal{K}_{A}^{x}$ as a function of frequency. Solid lines indicate the median of the data points $\langle \mathcal{K}_{A}^{x} \rangle$, dashed lines indicate $1\sigma$ error bars. Bottom: Histograms for $\mathcal{K}_{A}^{x}$, with the median shown in solid pink, and error bars in dashed pink. \label{asymmethod}}
\end{figure}

During the latter part of the science run, a more direct approach of determining the actuation coefficient $\mathcal{K}_{A}$ was taken, using the ``asymmetric Michelson technique.'' This method is similar in principle to the simple Michelson version of the free-swinging Michelson technique, however, we configure the interferometer as shown in Figure \ref{asymmich}. In this method the response of the end test mass is measured directly and
\begin{eqnarray}
\mathcal{K}_{A} & = & \left(\frac{q_{AS}}{exc_{e}}\right) \left(\frac{1 + G_{AM}}{k}\right) \left(\frac{1}{P_{cm}}\right) .
\end{eqnarray}
Figure \ref{asymmethod} shows an example result for L1.

\begin{figure}[h!]
\begin{center}
\includegraphics[width=90mm]{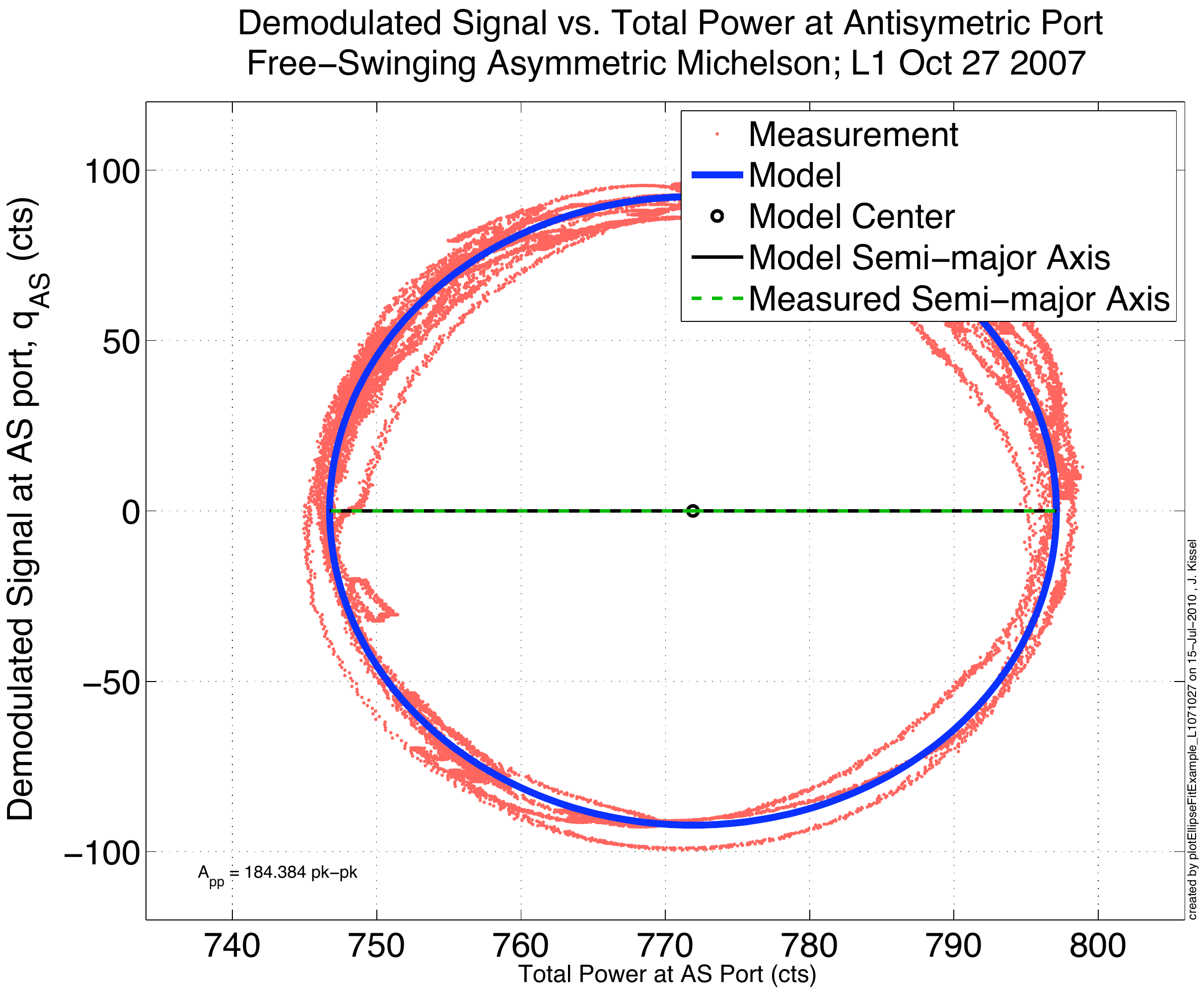}
\caption{An example ellipse produced by photodiode demodulated signal $q_{AS}$ versus total power, and corresponding fit to the ellipse used to determine $A_{pp}$ in asymmetric Michelson method. \label{Appcalc}}
\end{center}
\end{figure}

The quantity $k$ may vary slowly over the measurement period due to input laser power fluctuations, interferometer alignment, etc. The asymmetric Michelson is particularly sensitive to these variations as round trip power loss is large. For this method, we employ a more sophisticated technique for determining the amplitude $A_{pp}$, developed originally by Rolland et al \cite{virgocal}.  A plot of $q_{AS}$ versus the total power incident on the photodiodes should be an ellipse whose semi-minor axis is $A_{pp}/2$. We obtain a fit to this ellipse and extract $A_{pp}$ with a quantifiable statistical error.


\subsection*{Results}
\begin{figure}[h!]
  \begin{minipage}{70mm}
\includegraphics[width=70mm]{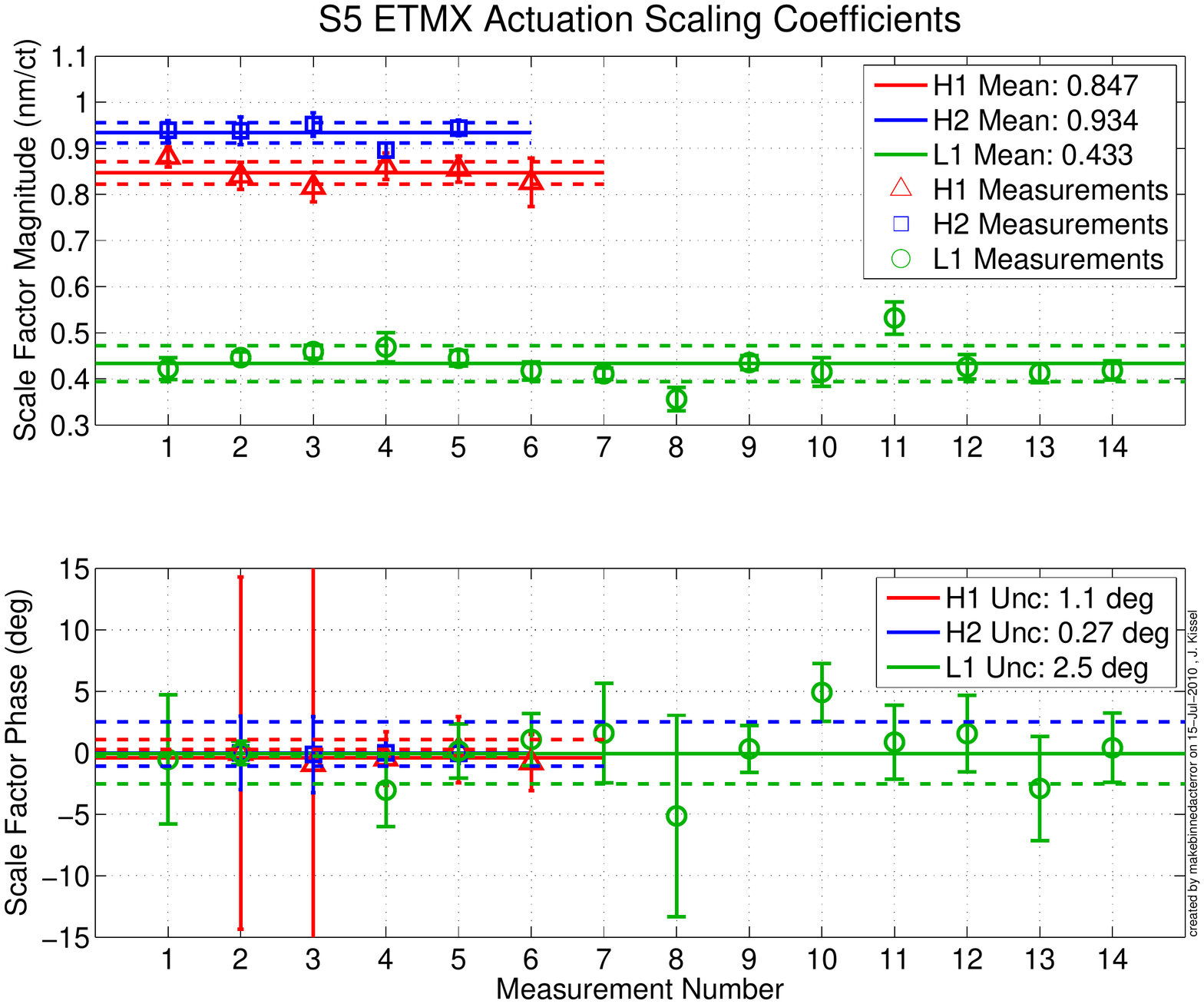}
  \end{minipage}
  \begin{minipage}{70mm}
\includegraphics[width=70mm]{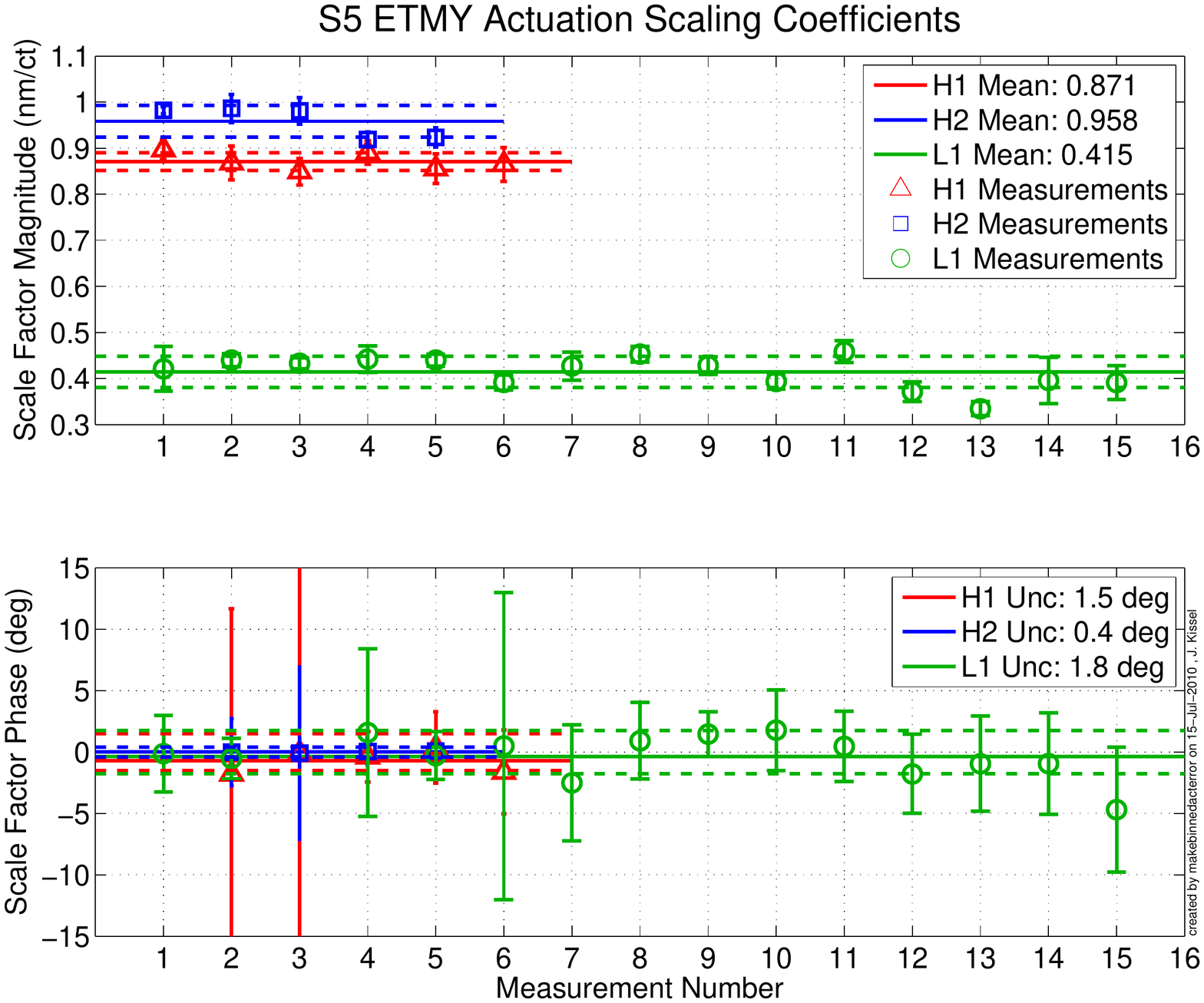}
  \end{minipage}
  \caption{Measurements of the actuation scaling coefficient $\mathcal{K}_{A}^{x,y}$, measured over the course of the fifth science run. Measurement numbers 6 in H1 and 7 through 15 in L1 used the asymmetric Michelson technique, the remainder were measured with the simple Michelson technique. (Left) Magnitude and phase of each measurement median $\langle \mathcal{K}_{A}^{x}\rangle_{j}$ (Top) and $\langle \mathcal{K}_{A}^{y} \rangle_{j}$ (Bottom) treated independently. The statistical uncertainty of actuation function is the quadrature sum of each arm's uncertainty, which takes the larger of the standard deviation of each measurements median, $\langle \mathcal{K}_{A} \rangle_{j}$ or the mean uncertainty divided by the number of measurements $\sigma_{\mathcal{K}_{A},j}$. \label{dccoeffs}}
  \end{figure}

Using the above methods, the actuation coefficient is measured many times for each optic in each interferometer over the course of the science run, and their mean used as the actuation scaling coefficient for all model epochs. Figure \ref{dccoeffs} shows the representative median and estimated uncertainty for each of these measurements. Table \ref{dccaltable} summarizes the actuation coefficients used in the actuation model, $\langle \mathcal{K}_{A} \rangle$ for the three interferometers in the fifth science run, using either simple Michelson or asymmetric Michelson techniques, with statistical uncertainty as described in \S\ref{acterror}.

\begin{table}[h!]
\caption{Summary of the actuation scaling coefficients measured during S5. These single numbers are formed by the mean of each measurement's median $\langle \mathcal{K}_{A} \rangle_{j}$ (6 for each end test mass in H1, 5 in H2, and 14 and 15 for the X and Y test masses, respectively in L1). Only statistical uncertainty is reported here; systematic uncertainty is folded the the total uncertainty of the actuation function. \label{dccaltable}}
\begin{center}
\begin{tabular}{|c|c|c|}
\hline
 & $  \mathcal{K}_{A}^{x} $ (nm/ct) & $  \mathcal{K}_{A}^{y} $ (nm/ct) \\
\hline
H1 & $0.847 \pm 0.024$ & $0.871 \pm 0.019$\\
H2 & $0.934 \pm 0.022$ & $0.958 \pm 0.034$\\
L1 & $0.434 \pm 0.039$ & $0.415 \pm 0.034$\\
\hline
\end{tabular}
\end{center}
\end{table}

Each simple Michelson measurement of a given optic's coefficient is assigned magnitude and phase uncertainty,
\begin{eqnarray}
\left(\frac{\sigma_{|\mathcal{K}_{A}|}}{|\mathcal{K}_{A}|}\right)^{2} & = & \left[\frac{\textrm{std}(|\mathcal{K}_{i}| )}{\langle |\mathcal{K}_{i}|\rangle} \right]^{2} +  \left[\frac{1}{\langle |\mathcal{R}_{ie}|\rangle}\frac{\textrm{std}(|\mathcal{R}_{ie}|)}{\sqrt{N}} \right]^{2} \\
\sigma_{\phi_{\mathcal{K}_{A}}}^{2} & = & \textrm{std}(\phi_{\mathcal{K}_{i}})^{2} + \left[\frac{\textrm{std}(\phi_{\mathcal{R}_{ie}})}{\sqrt{N}}\right]
\end{eqnarray}
and asymmetric Michelson measurement is assigned magnitude and phase uncertainty
\begin{eqnarray}
\left(\frac{\sigma_{|\mathcal{K}_{A}|}}{|\mathcal{K}_{A}|}\right)^{2} & = & \left[\frac{\textrm{std}(|\mathcal{K}_{A}| )}{\langle |\mathcal{K}_{A}|\rangle} \right]^{2}  \\
\sigma_{\phi_{\mathcal{K}_{A}}}^{2} & = & \textrm{std}(\phi_{\mathcal{K}_{A}})^{2}
\end{eqnarray}

In simple Michelson technique, measurements of $\mathcal{K}_{i}$  were found to be inconsistent with a Gaussian distribution across the frequency band. We therefore estimate the uncertainty in the median, $\langle \mathcal{K}_{i} \rangle$ to be the standard deviation alone. However, in the second step (Eq. \ref{simpmich2}), we have found the single arm transfer function ratio, $R_{ie}$, to be consistent with a Gaussian distribution across the frequency band, so we estimate the median uncertainty as though it were a gaussian distribution and divide the standard deviation by $\sqrt{N}$ where N is the number of frequency points. In the asymmetric Michelson method, where the measurement of $\mathcal{K}_{A}$ is similar to that of $\mathcal{K}_{i}$ in the simple Michelson method, we again do not assume a Gaussian distribution over the measurement band, and take the standard deviation alone.


\clearpage
\newpage
\section*{References}

\bibliographystyle{elsarticle-harv}
\bibliography{S5paper-NIMA.bib}







\end{document}